\documentclass[epj,nopacs]{svjour}
\usepackage{amsmath}         
\usepackage{amssymb}
\usepackage{mathtools} 
\usepackage{cuted} 
\usepackage[pdftex]{graphicx}	
\usepackage{grffile}
\usepackage[export]{adjustbox}
\usepackage{color}  				
\usepackage[pdftex]{hyperref} 
\usepackage[utf8]{inputenc}
\usepackage{dsfont}
\usepackage{slashed}
\usepackage{ulem}
\usepackage{widetext}

\numberwithin{equation}{section}

\newcommand{\ord}[1]{\mathcal{O}\left({#1}\right)}

\newcommand{\fig} [1]{Fig.~\ref{#1}}

\newcommand{\charm}{\tilde{\chi}^-}
\newcommand{\charp}{\tilde{\chi}^+}
\newcommand{\neut}{\tilde{\chi}^0}
\def\gsim{\raise0.3ex\hbox{$\;>$\kern-0.75em\raise-1.1ex\hbox{$\sim\;$}}}
\def\lsim{\raise0.3ex\hbox{$\;<$\kern-0.75em\raise-1.1ex\hbox{$\sim\;$}}}

\begin{document}

\title{Constraining sleptons at the LHC in a supersymmetric low-scale seesaw scenario}

\author{Nhell~Cerna-Velazco\inst{1}\thanks{n.cerna@pucp.edu.pe} \and Thomas~Faber\inst{2}\thanks{thomas.faber@physik.uni-wuerzburg.de} \and Joel~Jones-P\'erez\inst{1}\thanks{jones.j@pucp.edu.pe} \and Werner~Porod\inst{2}\thanks{porod@physik.uni-wuerzburg.de}}

\institute{Secci\'on F\'isica, Departamento de Ciencias, Pontificia Universidad Cat\'olica del Per\'u, Apartado 1761, Lima, Peru
\and
Institut f\"ur Theoretische Physik und Astrophysik, Uni W\"urzburg}

\date{Received: date / Revised version: date}

\abstract{
We consider a scenario inspired by natural supersymmetry, where neutrino data is explained within a low-scale seesaw scenario.
We extend the Minimal Supersymmetric Standard Model by adding light right-handed neutrinos and their superpartners,
the R-sneutrinos, and consider the lightest neutralinos to be higgsino-like.
\\
We consider the possibilities of having either an R-sneutrino or a higgsino as lightest supersymmetric particle.
Assuming that squarks and gauginos are heavy, we systematically
evaluate the bounds on slepton masses due to existing LHC data.
}

\maketitle

\section{Introduction}

The discovery of the Higgs boson in the 8 TeV run of the LHC
\cite{Aad:2012tfa,Chatrchyan:2012xdj} marks one of the most
important milestones in particle physics. Its mass is already known
rather precisely: $m_h = 125.09 \pm 0.21$~(stat.) $\pm
0.11$~(syst.)~GeV \cite{Aad:2015zhl}, and the signal strength of various LHC searches has been found
consistent with the SM predictions.
While this completes the Standard Model (SM) particle-wise, several
questions still remain open, for example: (i) Is it possible to include the SM
in a grand unified theory where all gauge forces unify? (ii)
Is there a particle physics explanation of the observed dark matter
relic density? (iii) What causes the hierarchy in the
fermion mass spectrum and why are neutrinos so much lighter than the
other fermions? What causes the observed mixing patterns in the fermion
sector? (iv) What stabilizes the Higgs mass at the electroweak scale?

Supersymmetric models address several of these questions and thus
the search for  supersymmetry (SUSY) is among the main priorities of
the LHC collaborations. Up to now no significant sign for physics beyond
SM has been found.  The
combination of the Higgs discovery with the (yet) unsuccessful
searches has led to the introduction of a model class called `natural
SUSY'
\cite{Brust:2011tb,Papucci:2011wy,Hall:2011aa,Blum:2012ii,Espinosa:2012in,D'Agnolo:2012mj,%
  Baer:2012uy,Younkin:2012ui,Kribs:2013lua,Hardy:2013ywa,Kowalska:2013ica,Han:2013kga}.
Here, the basic idea is to give electroweak-scale masses only to those SUSY particles 
giving a sizeable
contribution to the mass of the Higgs boson, such that a too
large tuning of parameters is avoided. All other particle masses are taken at the
multi-TeV scale. In particular, masses of the order of a few hundred GeV up to about one TeV are assigned to the higgsinos (the partners of the
Higgs bosons), the lightest stop (the partner of the top-quark) and, if the latter is mainly a left-stop, also to the light sbottom. 
In addition the gluino and the heavier stop masses should also be close to at most a few TeV.

Neutrino oscillation experiments confirm that at least two neutrinos have a non-zero mass. The exact mass generation mechanism for these particles is unknown, and both the SM and the MSSM remain agnostic on this topic. Although many ways to generate neutrino mass exist, perhaps the most popular one is the seesaw mechanism~\cite{Minkowski:1977sc,Yanagida:1979as,Mohapatra:1979ia,GellMann:1980vs,Schechter:1980gr,Foot:1988aq}. 
The main problem with 
the usual seesaw mechanisms
lies on the difficulty in testing its validity. In general, if Yukawa couplings are sizeable, the seesaw relations require Majorana neutrino masses to be very large, such that the new heavy states cannot be produced at colliders. In contrast, if one requires the masses to be light, then the Yukawas need to be small, making production cross-sections and decay rates to vanish. A possible way out of this dilemma lies on what is called the {\it inverse} seesaw~\cite{Mohapatra:1986bd}, which is based on having specific structures on the mass matrix (generally motivated by symmetry arguments) to generate small neutrino masses. This, at the same time, allows Yukawa couplings to be large, and sterile masses to be light.

We consider here a supersymmetric model where neutrino data are explained via a minimal inverse seesaw
scenario where the gauge-singlet neutrinos have masses in the range $\ord{{\rm keV}}$ to $\ord{100~{\rm GeV}}$. We explore this with a parametrization
built for the standard seesaw, and go to the limit where the inverse seesaw emerges, such that Yukawas and mixings become sizeable.
Although non-SUSY versions of this scenario can solve the dark matter and matter-antimatter asymmetry problems~\cite{Dodelson:1993je,Akhmedov:1998qx,Asaka:2005pn}, we shall make no claim on these issues in our model.

In view of the naturalness arguments, we also assume that the higgsinos have masses of $\ord{100~{\rm GeV}}$, whereas the gaugino masses lie
at the multi-TeV scale (see~\cite{Carpenter:2016lgo} for an example of such a scenario).
In addition, we assume all squarks are heavy enough such that LHC bounds are avoided, and play no role in the phenomenology within this work\footnote{Note, that even a light stop with mass of 3 TeV is consistent with 3\% fine-tuning
in the context of high scale models with non-universal Higgs mass parameters, see e.g.~\cite{Baer:2017yqq}
and refs.\ therein.}. In contrast we allow for fairly light sleptons and investigate the extent to which
current LHC data can constrain such scenarios. For further studies in other regions of the parameter space, see~\cite{Deppisch:2004fa,Bazzocchi:2009kc,Hirsch:2009ra}.

This paper is organized as follows: in the next section we present the model. Section \ref{sec:scan} summarizes
the numerical tools used and gives an overview of the LHC analysis used for these investigations. In Section 
\ref{sec:results} we present our findings for the two generic scenarios which differ in
the nature of the lightest supersymmetric particle (LSP): a higgsino LSP and a sneutrino LSP.
In Section \ref{sec:conclusion} we draw our conclusions. Appendices \ref{app:numass} and \ref{app:snumassmatrix}
give the complete formulae for the neutrino and sneutrino masses.

\section{The Model}
\label{sec:model}

We add three sterile neutrino superfields $\hat\nu_{R,k}$, and assume conserved $R$-parity. With this, the superpotential reads as 
\begin{align}
\mathcal{W}_{eff} = \mathcal{W}_{\rm MSSM}
+ \frac{1}{2} (M_R)_{ij}\,\hat{\nu}_{R,i}\,\hat{\nu}_{R,j}
+ (Y_\nu)_{ij}\,\widehat{L}_i \cdot \widehat{H}_u\, \hat{\nu}_{R,j}
\end{align}

The corresponding soft SUSY breaking terms are given by
\begin{eqnarray}\mathcal{V}^{soft} &=&\mathcal{V}_{\rm MSSM}^{soft}
  + (m^2_{\tilde\nu_R})_{ij}\tilde{\nu}^*_{R,i}\tilde{\nu}_{R,j}
  + \frac{1}{2}(B_{\tilde\nu})_{ij}\tilde{\nu}_{R,i}\tilde{\nu}_{R,j} \nonumber \\ 
  &&+ (T_\nu)_{ij}\,\tilde{L}_i \cdot H_u\, \tilde{\nu}_{R,j}
\end{eqnarray}

For the neutrino sector we use a Casa-Ibarra-like pa-rametrization~\cite{Casas:2001sr,Donini:2012tt}, the details of which can be found in Appendix~\ref{app:numass}. In this work, for simplicity, we shall use a non-trivial $R$ matrix which will enhance the Yukawa couplings of the two heaviest neutrinos, allowing us to write:
\begin{subequations}
\label{eq:YukawasSimple}
\begin{eqnarray}
 (Y_\nu)_{a 5} &=& \pm (Z^{\rm NH}_a)^*\sqrt{\frac{2m_3 M_5}{v_u^2}}\cosh\gamma_{56}\,e^{\mp i\rho_{56}}~, \\
 (Y_\nu)_{a 6} &=& -i (Z^{\rm NH}_a)^*\sqrt{\frac{2m_3 M_6}{v_u^2}}\cosh\gamma_{56}\,e^{\mp i\rho_{56}}~.
\end{eqnarray}
\end{subequations}
Here, $m_3$ is the mass of the largest light neutrino mass, $M_i$ are the masses of the heaviest neutrinos, and $a=e,\,\mu,\,\tau$. The parameters $\rho_{56}$ and $\gamma_{56}$ are the real and imaginary components of a complex mixing angle, appearing in the $R$ matrix. The $Z_a^{\rm NH}$ factors~\cite{Gago:2015vma} depend on the PMNS mixing matrix and ratios of light neutrino masses, and are in general of $\ord{1}$. The only exception is $Z_{e}^{\rm NH}$, which is slightly suppressed due to the small $s_{13}$.

The Yukawas can be significantly enhanced by taking a large $\gamma_{56}$. 
Furthermore, we can see that, if $M_5=M_6$, the two Yukawa couplings have the same size. From here, it is straightforward to 
redefine the sterile states, and demonstrate that the resulting mass matrix has the same structure as the one 
of the inverse seesaw.

In this work, we denote $\nu_L=\nu_{1,2,3}$ and $\nu_h=\nu_{5,6}$. Since the Yukawas for the lightest right-handed neutrino $\nu_4$ are not enhanced, this particle effectively decouples in the model. For definiteness, we take the neutrino oscillation parameters $s_{12}^2=0.304$, $s^2_{13}=0.0218$, $s^2_{23}=0.452$, $\Delta m^2_{21}=7.5\times10^{-5}$~eV$^2$, $\Delta m^2_{31}=2.5\times10^{-3}$~eV$^2$, with all CP phases equal to zero. For the heavy neutrino sector, we set $M_4=7$~keV, $M_5=M_6=20$~GeV and $\gamma_{56}=8$. The latter choice is taken such that the non-SUSY contribution does not saturate lepton flavour violation (LFV) bounds~\cite{Gago:2015vma}. With these values, the largest neutrino Yukawa coupling becomes of $\ord{10^{-4}}$.

For the sneutrino sector, we have written the full sneutrino mass matrix in Appendix~\ref{app:snumassmatrix}. For simplicity, we neglect terms proportional to $Y_\nu$, and take vanishing $B_{\tilde\nu}$ and $T_\nu$\footnote{We have checked that this is a very good approximation if $B_{\tilde\nu}\leq10^{-4}\times(m_{\tilde \nu}^2+M_R^\dagger M_R)$ as well as having $T_\nu\sim\ord{Y_\nu\times10~{\rm TeV}}$.}. In this case, we do not need to split the sneutrino fields into scalar and pseudoscalar components, and can work with the $\tilde \nu_L$ and $\tilde\nu_R^c$ states. We can then approximately write the sneutrino mass matrix as:
\begin{equation}
 M_{\tilde \nu}^2=\left(\begin{array}{cc}
m_{\tilde L}^2+\frac{1}{2}m_Z^2\cos2\beta & 0 \\
0 & 
m_{\tilde \nu_R}^2+M_R^\dagger M_R
\end{array}\right) 
\end{equation}
such that we can assume that three $\tilde\nu_i$ states shall be dominantly $\tilde\nu_L$, and other three states shall be 
dominantly $\tilde\nu_R$. Thus, we refer to them as L-sneutrinos and R-sneutrinos, respectively. 
In the following, we take $m^2_{\tilde L}$ and $m^2_{\tilde\nu_R}$ flavour diagonal, so the only 
source of sneutrino mixing comes from $M_R$, which is very small. Thus, we denote the L-sneutrinos through 
their interaction eigenstates ($\tilde\nu_{e L}$, $\tilde\nu_{\mu L}$, $\tilde\nu_{\tau L}$), while R-sneutrinos
are denoted as $\tilde\nu_{1,2,3}$. Notice that in our results we use the full formulae shown in Appendices~\ref{app:numass} and~\ref{app:snumassmatrix}.

As usual, the model contains neutralinos and chargi-nos. As mentioned before, we assume that the 
gaugino mass parameters
are much larger than the higgsino mass parameter $|\mu|$. Therefore, the lightest states are two higgsi-no-like neutralinos
$\neut_{1,2}$ and a higgsino-like chargino $\charm$ which are nearly mass degenerate,
see e.g.\ \cite{Barducci:2015ffa} for a discussion of the resulting spectrum. 

The best way to probe this model and to distinguish it from the MSSM is by discovering and 
studying the R-(s)neutrino properties. However, it is clear 
that their direct production at the LHC is not a very efficient process, as the cross sections
are proportional to the corresponding Yukawa couplings. A better way to generate them is through cascade 
decays of heavier particles, such as L-sleptons or higgsinos. Thus, in this work, we always consider 
L-sneutrinos heavier than R-sneutrinos.

\begin{figure}[tbp]
\centering
\includegraphics[width=0.48\textwidth]{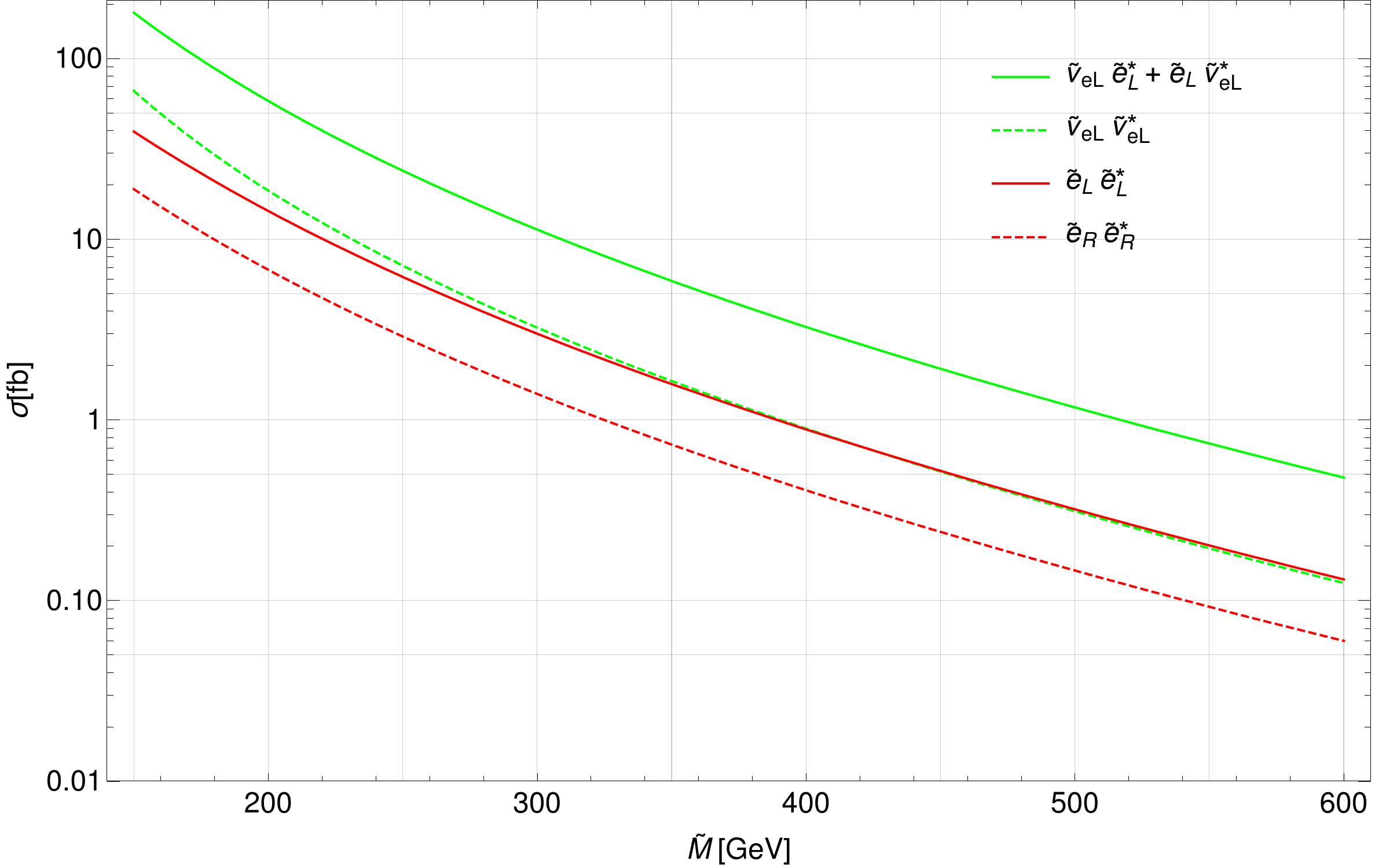}
\caption{Various tree-level 
cross sections in fb for the production of one generation of sleptons and sneutrinos at the LHC with 13 TeV
as a function of the corresponding soft SUSY mass parameter $\tilde M$: green (bright) full line 
$\sigma(pp\to \tilde e_L  \tilde \nu_{eL}^*) + \sigma(pp\to \tilde e_L^*  \tilde \nu_{eL})$, 
green (bright) dashed line $\sigma(pp\to  \tilde \nu_{eL} \tilde \nu_{eL}^*)$,
red (dark) full line $\sigma(pp\to \tilde e_L  \tilde e_L^*)$ and
red (dark) dashed line $\sigma(pp\to \tilde e_R  \tilde e_R^*)$. $\tilde M$ is either the soft SUSY breaking
parameter 
$M_{\tilde L}$ or $M_{\tilde E}$ depending on the particles considered.}
\label{fig:Xsections}
\end{figure}
In Figure~\ref{fig:Xsections} we present the cross sections for $\tilde e_{L,R}$/$\tilde\nu_L$ production
at tree-level. Notice that these cross sections are the same for the $\tilde \mu_{L,R}$/$\tilde\nu_{\mu L}$ and 
$\tilde \tau_{L,R}$/$\tilde\nu_{\tau L}$ flavours.
It is well-known that QCD corrections shift these to larger values
\cite{Fuks:2013vua}, so we apply an overall K-factor of 1.17. Note that the sum of the processes
$pp\to \tilde e_L  \tilde \nu_{e L}^*$ and $pp\to \tilde e_L^*  \tilde \nu_{e L}$ has by far
the largest cross section, followed by $\tilde e_L\tilde e_L^*$ and $\tilde\nu_{e L}\tilde\nu_{e L}^*$ pair production.
In the following, we focus on the resulting signal from the decay of these states, as they
explain the main features of our results. Nevertheless, in the numerical analysis we have included
all possible processes, such that the available data is fully exploited.

Note that left-right mixing in the stau sector is large, such that both $\tilde\tau^-_1$ and $\tilde\tau^-_2$ have a relatively 
large $\tilde\tau^-_L$ component. This means that in the following we need to study the decays of both states.

The final states, and thus the signal, depend on the nature of the LSP, which can be either an R-sneutrino or a neutralino. 
Moreover, in case of  an  R-sneutrino  LSP, we also have a different 
phenomenology depending on wheth-er the higgsinos are lighter or heavier than the L-sleptons. In 
addition, in the case of  higgsinos being lighter than the L-sleptons, we shall also have a 
dependence on the size of the small gaugino admixture to the physical charginos and neutralinos.

Before we give details of each scenario, we first review the relevant part of the interaction Lagrangian of $\tilde l_L$ and $\tilde l_R$ with
 charginos and neutralinos:
\begin{align}
{\cal L} = \sum_{\substack{i=1,\dots,4\\j=L,R}}
\big{(}&\bar{l} ( c^L_{ij\tilde{l}} P_L + c^R_{ij\tilde{l}} P_R) \neut_i \tilde{l}_j \nonumber \\
 &+\bar{\nu} ( c^L_{ij\tilde{\nu}} P_L + c^R_{ij\tilde{\nu}} P_R) \neut_i \tilde{\nu}_j + h.c. \big{)} \nonumber \\
+\sum_{\substack{k=1,2\\j=L,R}} \big{(}
&( \bar{\nu}_R d^L_{kj\tilde{l}} P_L +  \bar{\nu}_L d^R_{kj\tilde{l}} P_R ) \charp_k  \tilde{l}_j \nonumber \\
&+\left( \bar{l}_R d^L_{kj\tilde{\nu}} P_L +  \bar{l}_L d^R_{kj\tilde{\nu}} P_R \right)\charp_k  \tilde{\nu}_j 
+ h.c. \big{)}
\end{align}
with 
\begin{align}
c^L_{iL\tilde{l}} &= - Y_l N_{i3}^*  &    c^R_{iL\tilde{l}} &= \frac{1}{\sqrt{2}} \left( g' N_{i1} + g N_{i2} \right) \\
c^L_{iR\tilde{l}} &= -\sqrt{2}  g' N_{i1}^*  &  c^R_{iR\tilde{l}} &=  - Y_l N_{i3}\\
c^L_{iL\tilde{\nu}} &= - Y_\nu N_{i4}^* &    c^R_{iL\tilde{\nu}} &= \frac{1}{\sqrt{2}} \left( g' N_{i1} - g N_{i2} \right) \\
c^L_{iR\tilde{\nu}} &= 0  &  c^R_{iR\tilde{\nu}} &=  - Y_\nu N_{i4}
\end{align}
\begin{align}
d^L_{kL\tilde{l}} &= Y_\nu V_{k2}^* &    d^R_{kL\tilde{l}} &= - g U_{k1}  \\
d^L_{kR\tilde{l}} &= 0  &    d^R_{kR\tilde{l}} &=   Y_l U_{k2}  \\
d^L_{kL\tilde{\nu}} &=  Y_l U_{k2}^* &    d^R_{kL\tilde{\nu}} &= - g V_{k1}  \\
d^L_{kR\tilde{\nu}} &= 0  &    d^R_{kR\tilde{\nu}} &=   Y_\nu V_{k2} 
\end{align}
where, for simplicity, we have neglected generation indices as well as left-right mixing. This is a very good approximation
for the sneutrinos, the first two slepton generations, and for the staus in case of small to medium values of $\tan\beta$.

The neutralino mixing matrix $N$ is in the basis $\tilde b$, $\tilde w^3$, $\tilde H_d$, $\tilde H_u$, and in our model
we have $|N_{i1}|,|N_{i2}| \ll |N_{i3}|,|N_{i4}|$, for $i=1,2$. Moreover, $U$ and $V$ are the 
chargi-no mixing matrices, in the basis $\tilde w^\pm,\tilde H^\pm$, such that in our model we have
$|U_{11}|,|V_{11}| \ll |U_{12}|,|V_{12}|$. In addition, we know  that $Y_l,Y_\nu \ll g',g$, the only exception being
$Y_\tau$, which can become $\ord{g'}$ in case of very large $\tan\beta$.

Knowing these couplings is very convenient at the time of understanding the different branching ratios. For example, if we want to compare the $\tilde \mu_L$ decays into higgsinos, for very large values of $M_1$, $M_2$, we find:
\begin{equation}
 \frac{{\rm BR}(\tilde\mu_L^-\to\nu_L\tilde\chi^-_k)}{{\rm BR}(\tilde\mu_L^-\to\mu^-\tilde\chi^0_i)}\sim\frac{|g\, U_{k1}|^2}{|Y_\mu N_{i3}|^2}
\end{equation}
where we see that decays into charginos will be subdominant if the mixing with gauginos is small enough.

\subsection{Sneutrino LSP and light higgsinos ($m_{\tilde\nu_R}<\mu<m_{\tilde L}$)}
\label{sec:snuLSP.mulight}

This scenario is characterized by subsequent two-body decays. The heavier L-sleptons decay into states involving $\tilde\chi^0_{1,2}$ or $\tilde\chi^\pm_1$, which then decay into states involving R-sneutrinos. The decay chains have several branches, with the dominant branching ratio for L-sleptons depending on the size of the couplings and the respective elements of neutralino and chargino mixing matrices.

In the following, for each slepton, we compare two scenarios. In the first one, we set $M_1=M_2=2$~TeV, such that there is a small but non-negligible gaugino admixture on the neutralino and chargino states. On the second one, both gauginos are ``decoupled'' from the model by setting their masses at 1 PeV.

\begin{figure*}
\centering
\includegraphics[scale=0.5]{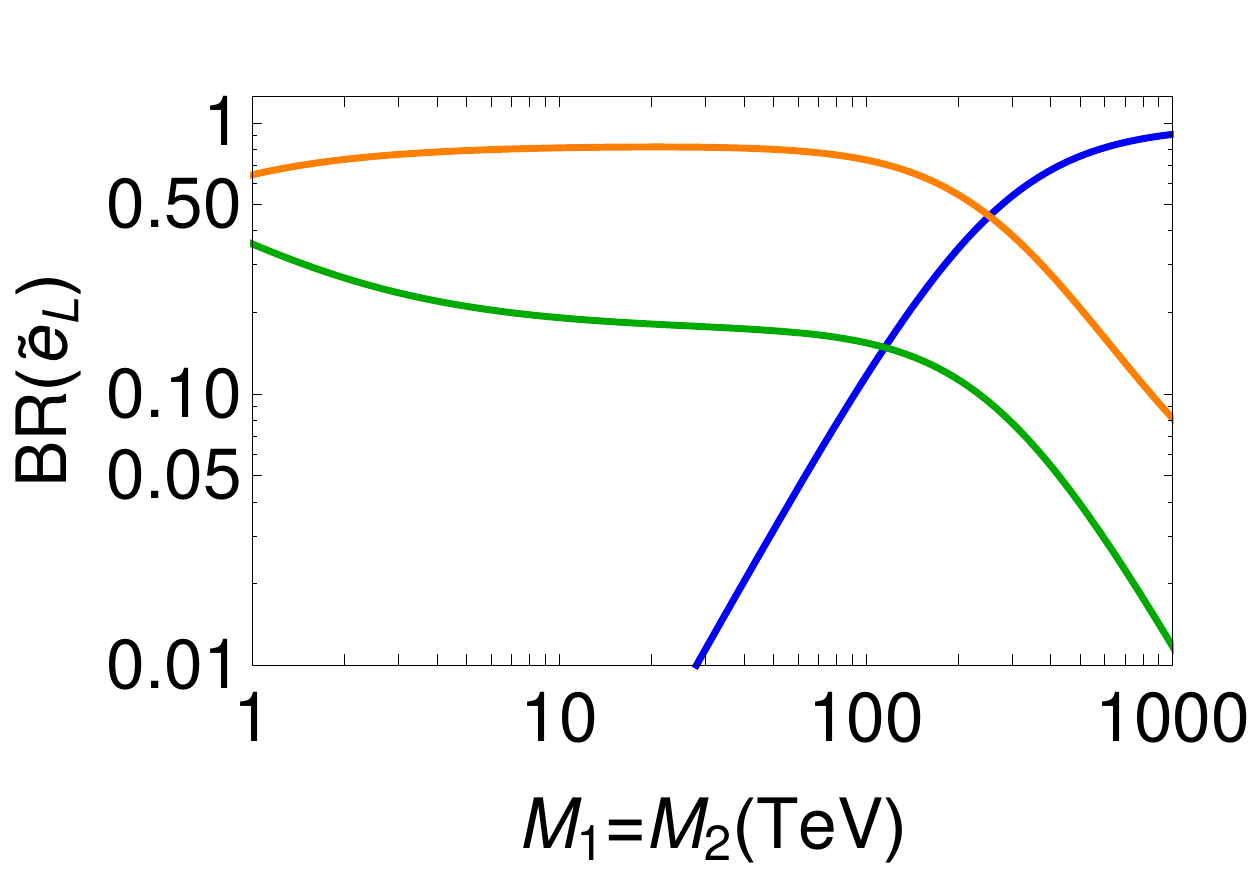} \quad
\includegraphics[scale=0.5]{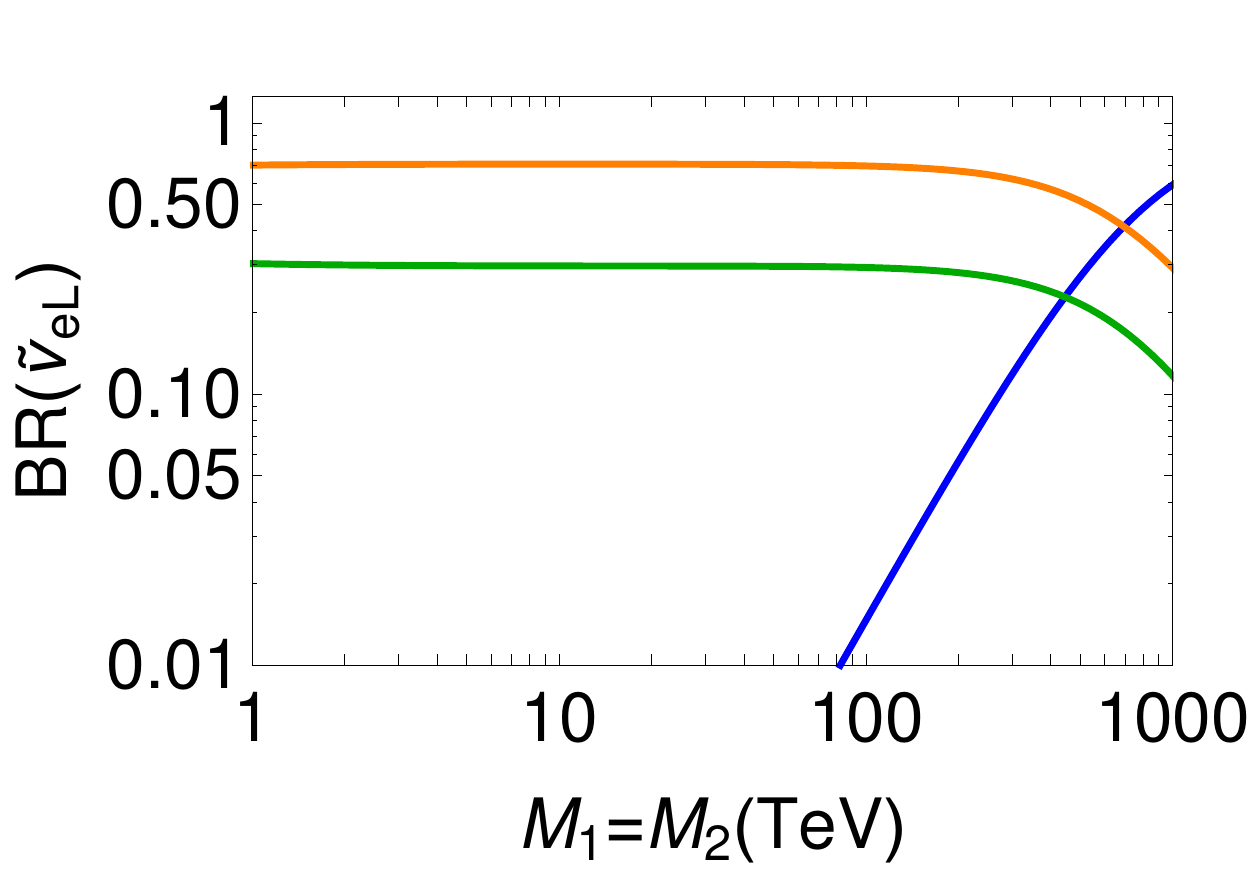} \\
\includegraphics[scale=0.5]{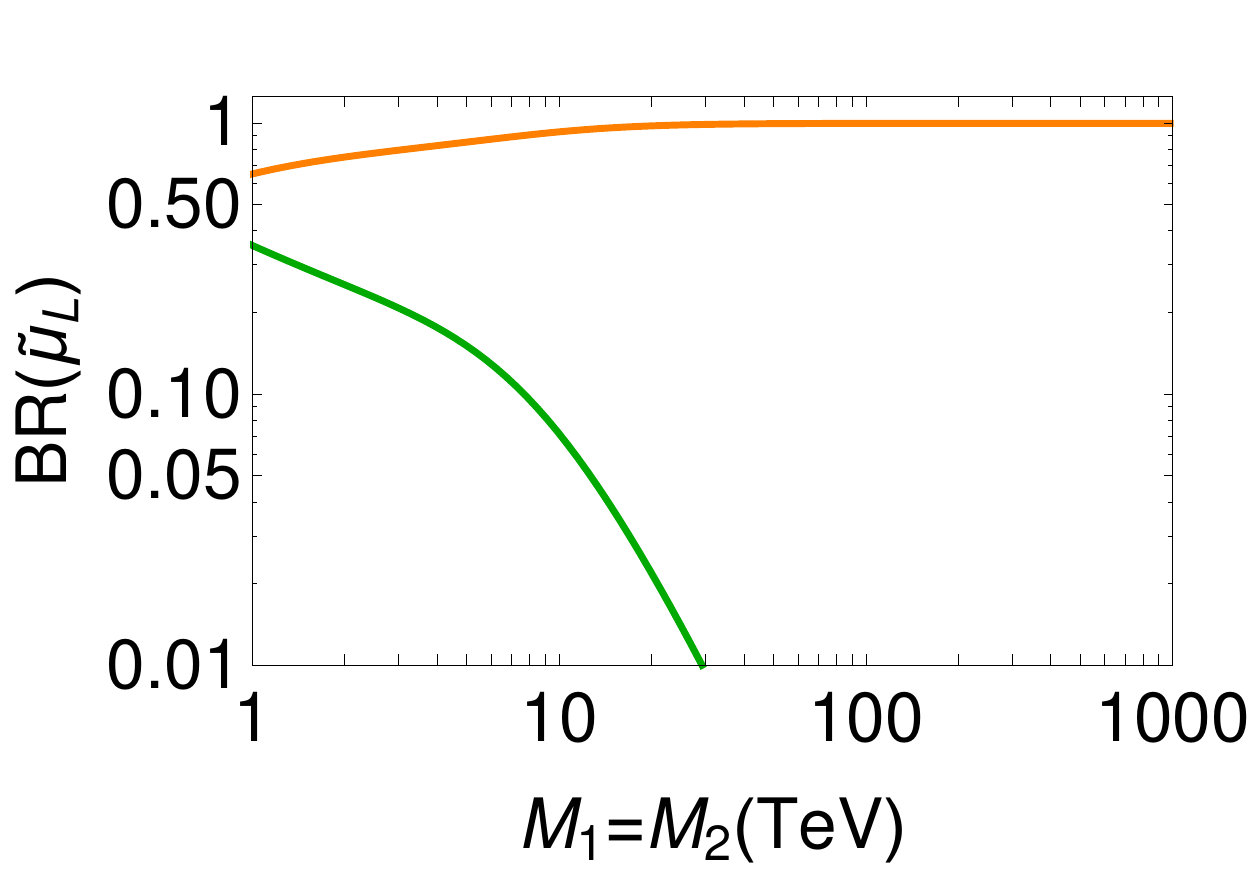} \quad
\includegraphics[scale=0.5]{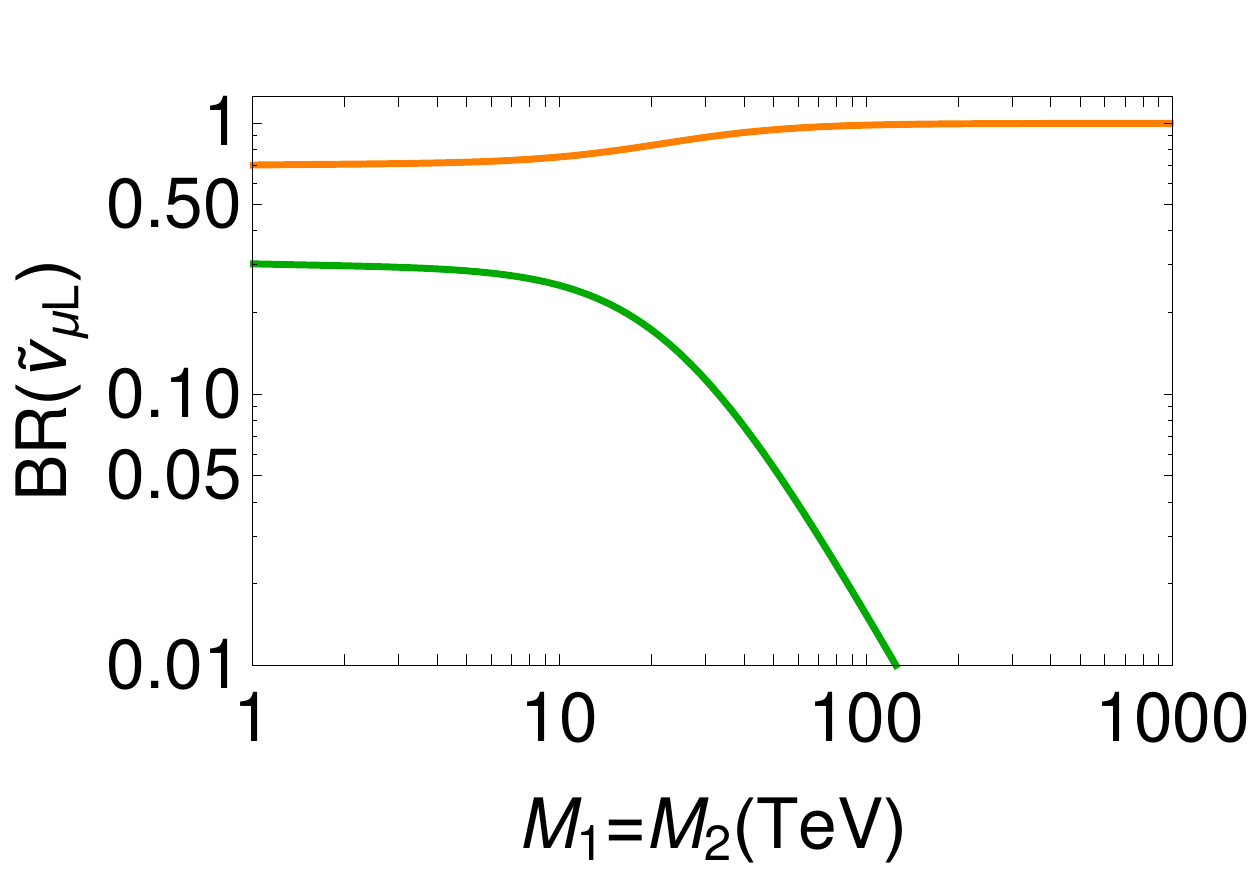} \\
\includegraphics[scale=0.5]{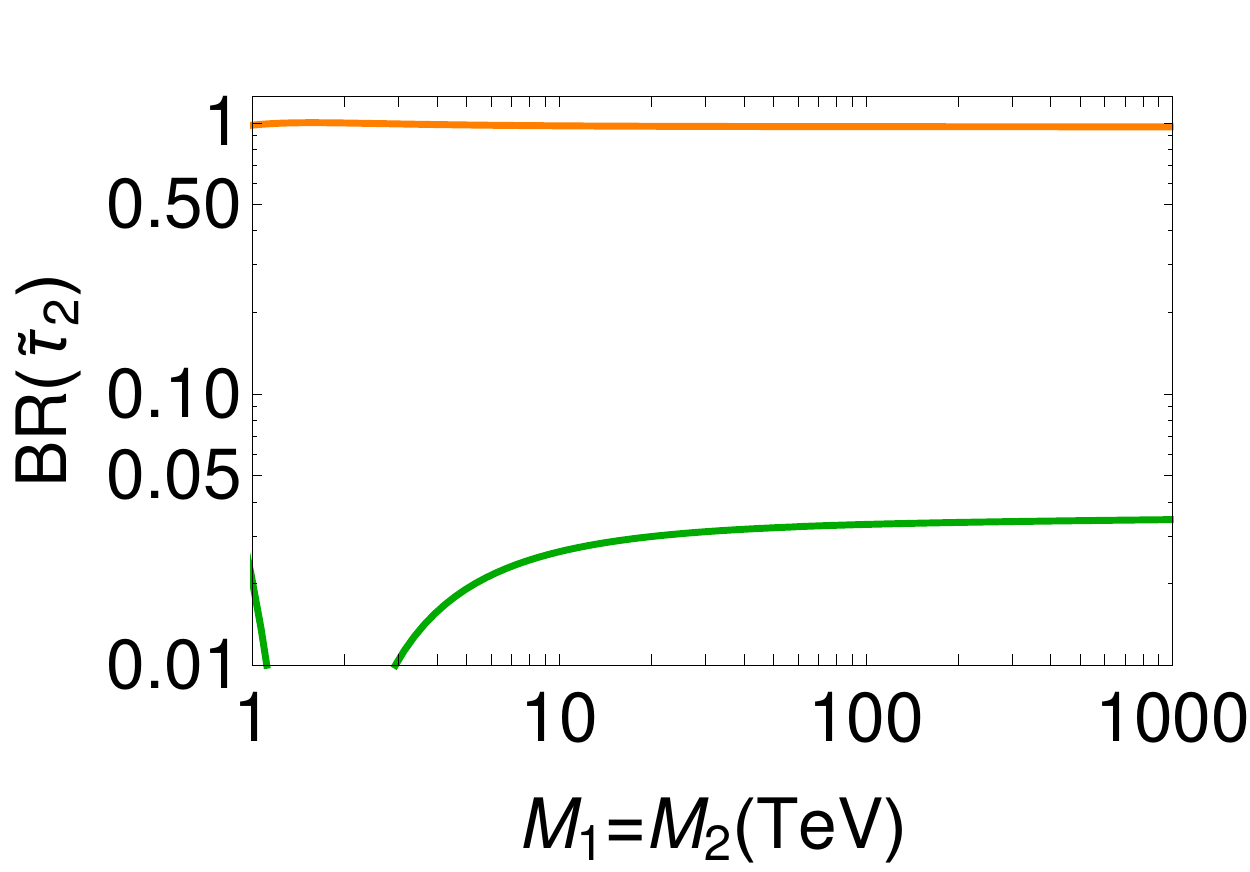} \quad
\includegraphics[scale=0.7]{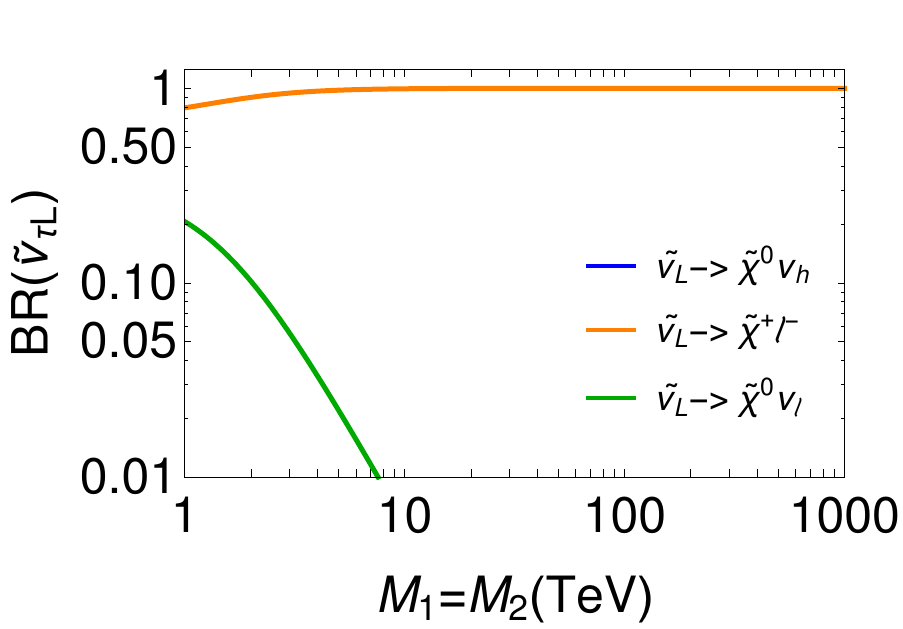} \\
\includegraphics[scale=0.7]{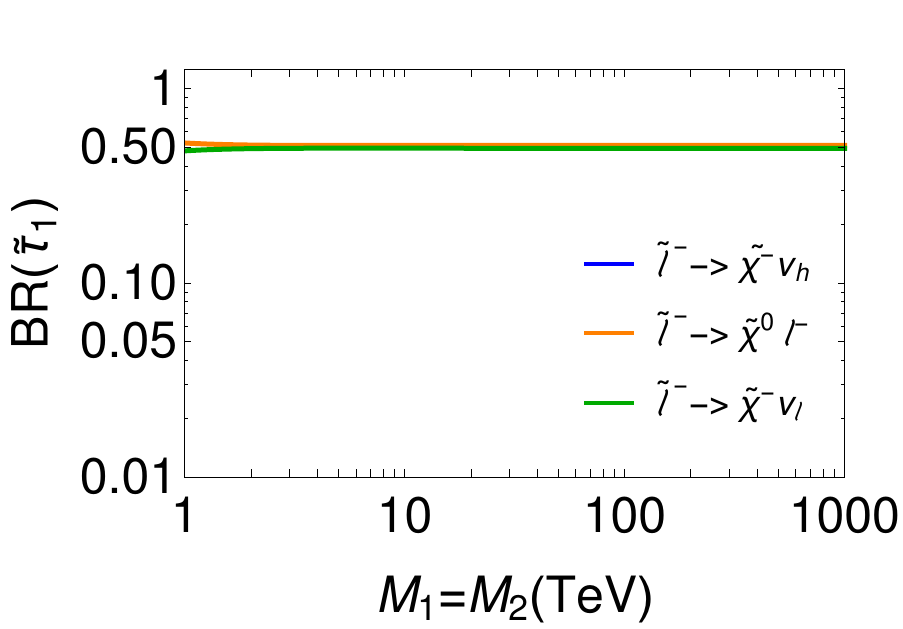} \quad \hspace{177pt}
\caption{\label{fig:SleptonBR1} Branching ratios for sleptons as a function of gaugino mass $M_1=M_2$, for $\mu<m_{\tilde L}$ Decays for charged sleptons (sneutrinos) are shown on the left (right) column. The last panels describe the colours for the branching ratios shown in each column. In case of neutralinos, the sum over the two lightest states is shown.}
\end{figure*}

For definiteness, we shall set $m_{\tilde L}=600~{\rm GeV}$, $\mu=120$~GeV and $\tan\beta=6$. In Figure~\ref{fig:SleptonBR1}, we show the most important branching ratios of each slepton as a function of $M_1=M_2$, which shall now be discussed. Notice we neglect to comment those cases where two contributions interfere destructively, as this effect is not of our interest.

For the smuon $\tilde\mu_L^-$, the 2 TeV gaugino scenario leads to primarily 
$\tilde\mu^-_L\to\mu^-\neut$ ($75\%$), followed by $\tilde\mu^-_L\to\nu_L\tilde\chi^-$ ($25\%$). 
The latter decay is due to a gauge coupling, and its branching ratio vanishes when the gauginos decouple. 
In contrast, $\tilde\mu^-_L\to\mu^-\neut$ is due to a combination of $Y_\mu$ and gauge contributions, and its branching ratio 
rises to unity in the decoupling regime. The muon sneutrino $\tilde\nu_{\mu L}$ follows a very similar pattern, 
with $\tilde\nu_{\mu L}\to\mu^-\charp$ dominating ($70\%$ at 2 TeV, and then $100\%$ in the decoupling scenario), 
followed by $\tilde\nu_{\mu L}\to\nu_L\neut$, which decreases in front of rising gaugino masses.

The case for the selectron $\tilde e_L^-$ is very similar to the one for $\tilde\mu_L^-$ in the 2 TeV case, replacing $\mu$ 
by $e$, and with very similar branching ratios. However, in the gaugino decoupling scenario, we find that the most 
relevant decay is $\tilde e_L^-\to\nu_h\charm$ ($90\%$), followed by $\tilde e_L^-\to e^-\neut$ ($9\%$) 
and $\tilde e_L^-\to \nu_L\charm$ ($1\%$). The reason for this is that the first decay proceeds through a 
$Y_\nu$ coupling, which is larger than the $Y_e$ coupling that governs the second decay. 
Again, the electron sneutrino $\tilde\nu_{e L}$ decays are similar to the $\tilde\nu_{\mu L}$ for 2 TeV gauginos, and 
in the decoupling case they change to $\tilde\nu_{e L}\to\nu_h\neut$ ($60\%$), $\tilde\nu_{e L}\to e^-\charp$ ($30\%$) 
and $\tilde\nu_{e L}\to\nu_L\neut$ ($10\%$).

Stau decays are somewhat unique, as the mass eigenstates have large components of both $\tilde\tau_L^-$ 
and $\tilde\tau_R^-$. We find that the lightest stau, $\tilde\tau_1^-$, which is mostly $\tilde\tau_R^-$, decays in 
equal proportions through $\tilde\tau_1^-\to\tau^-\neut$ and $\tilde\tau^-_1\to\nu_L\tilde\chi^-$ ($50\%$). In contrast, the 
heaviest stau, mainly $\tilde\tau_L$, decays through $\tilde\tau^-_2\to\tau^-\neut$ ($97\%$) and 
$\tilde\tau^-_2\to\nu_L\charp$ ($3\%$). Ignoring interference effects, the branching ratios of both $\tilde\tau_1$ and $\tilde\tau_2$ are independent of the mass of the gauginos. The reason for this is that the stau states can always couple with higgsinos through the $Y_\tau$ coupling, which is relatively large. For $\tilde\tau_2$, the difference in the values 
of the branching ratios is due to $\tilde\tau_2^-\to\nu_L\charp$ being somewhat suppressed due to the need of $LR$ mixing.

The tau sneutrino $\tilde\nu_{\tau L}$ follows a different pattern, as here there is no large mixing with any $\tilde\nu_R$, such that gaugino couplings can play an important role again. 
For the 2 TeV gaugino scenario, the dominating decay is $\tilde\nu_{\tau L}\to\tau^-\charp$ ($90\%$), followed by 
$\tilde\nu_{\tau L}\to\nu_L\neut$ ($10\%$). The former increases to ($100\%$) in the decoupled case, as in the $\tilde\nu_{\mu L}$ scenario.

In all scenarios, the charginos decay into a charged lepton, and a light sneutrino: $\charm_1\to\ell^-\tilde\nu_{1,2,3}$. 
The charged lepton is usually a muon or a tau, due to the $Z^{\rm NH}_a$ factors in the Yukawa couplings, Eqs~(\ref{eq:YukawasSimple}), 
with the branching ratio into an electron being below 10\%.
Moreover, due to their couplings, the branching ratio of the decay into $\tilde\nu_1$ is very suppressed, 
so charginos decay mostly into $\tilde\nu_{2,3}$. In principle, these should decay further through 3-body 
processes into additional leptons and $\tilde\nu_1$. However, we find $\tilde\nu_{2,3}$ to be long-lived, 
and escape the detector. Thus, charginos contribute to our signal with a charged lepton and missing energy.
Neutralinos follow a similar trend, but decay into a light neutrino instead of a charged lepton. 
Thus, neutralinos can be considered missing energy.

Finally, one has to take into account the decays of the heavy neutrinos, which are independent of the 
SUSY scenario considered. The heaviest neutrinos form a pseudo-Dirac pair, and shall decay 
promptly~\cite{Atre:2009rg}. We shall concentrate on decays involving at least one charged lepton: 
$\nu_{5,6}\to\ell^-qq'$ or $\nu_{5,6}\to\ell^{-(\prime)}\ell^+\nu_{\ell}$, with off-shell mediators. 

\subsection{Sneutrino LSP and heavy higgsinos ($m_{\tilde\nu_R}<m_{\tilde L}<\mu$)}
\label{sec:snuLSP.muheavy}

\begin{figure*}
\centering
\includegraphics[scale=0.5]{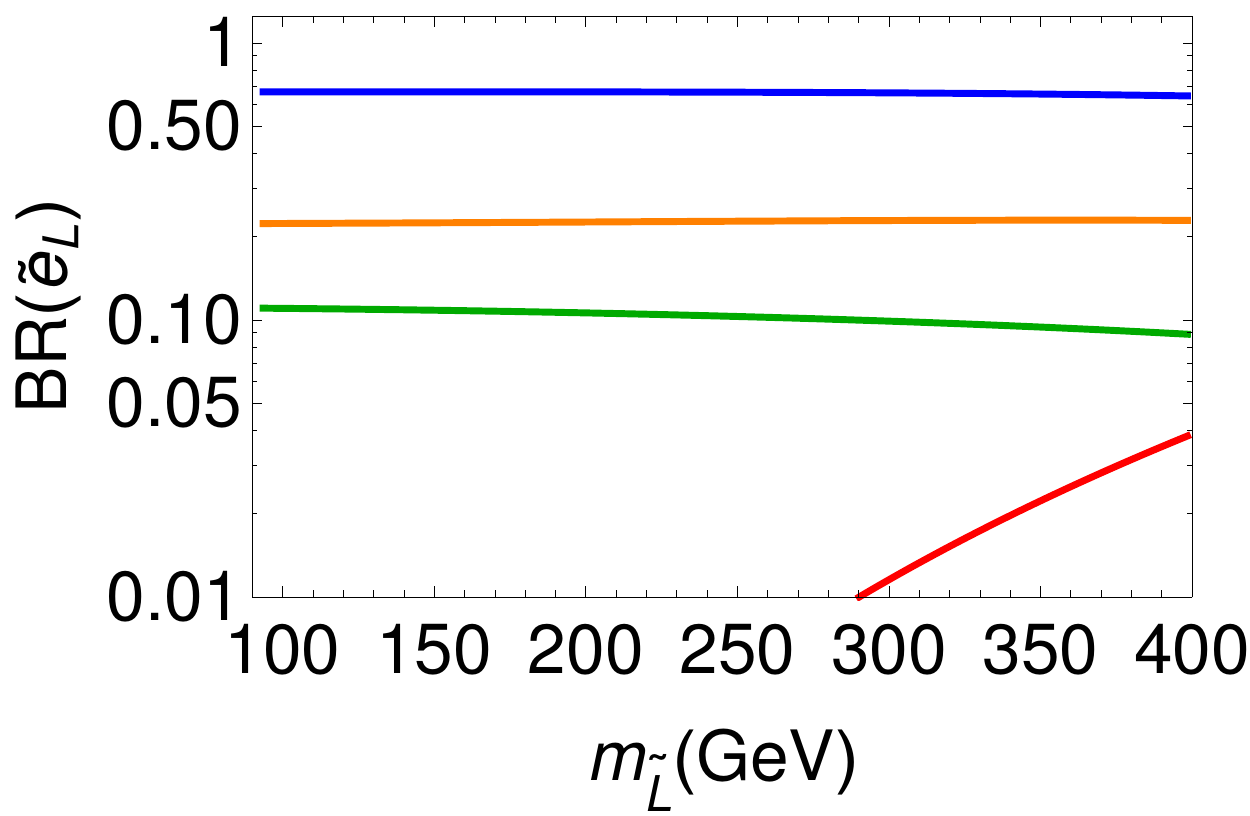} \quad
\includegraphics[scale=0.5]{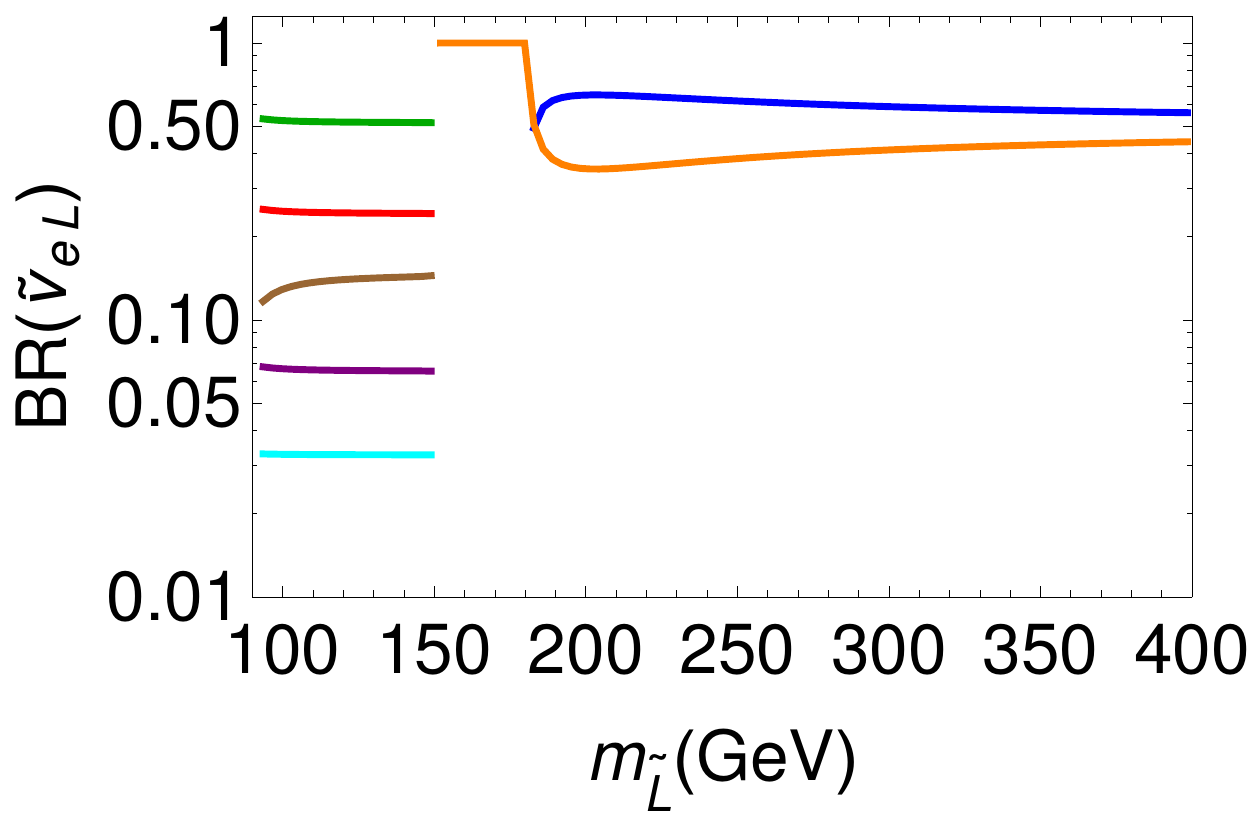} \\
\includegraphics[scale=0.5]{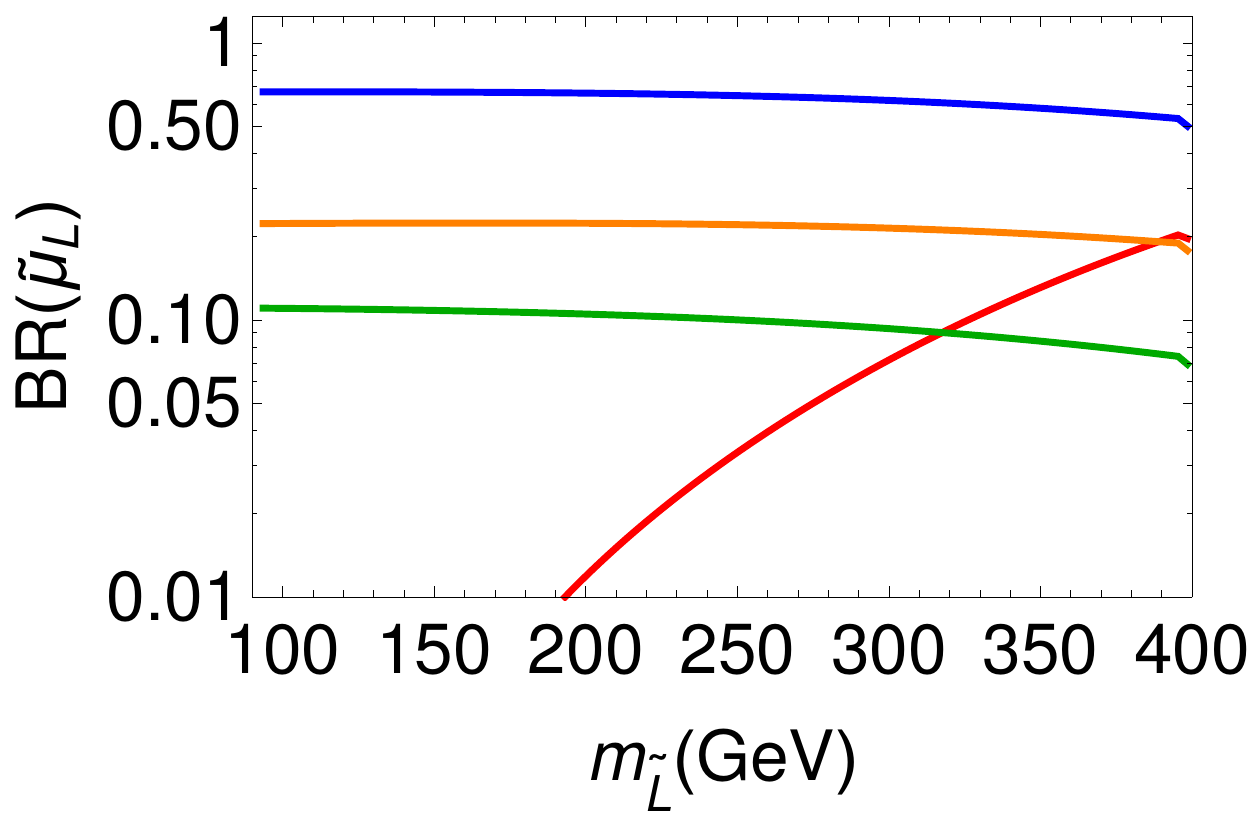} \quad
\includegraphics[scale=0.5]{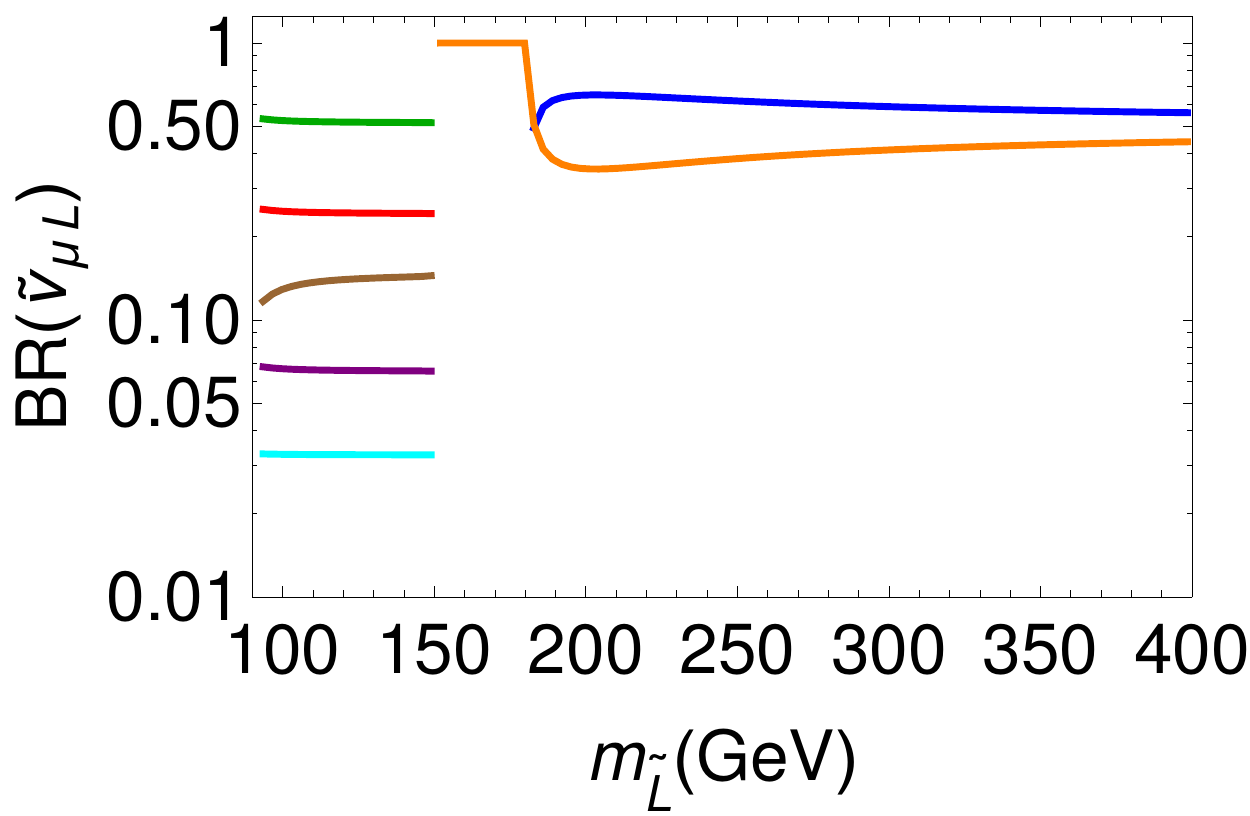} \\
\includegraphics[scale=0.5]{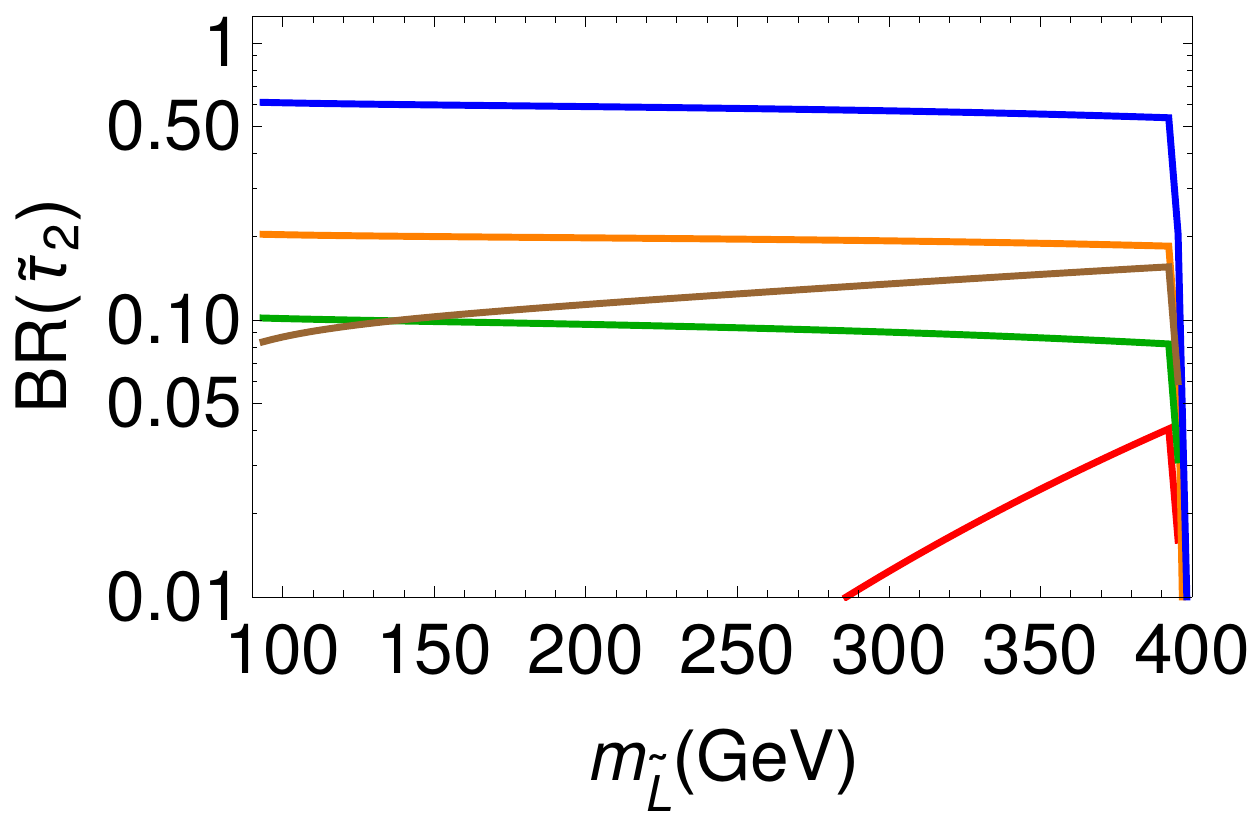} \quad
\includegraphics[scale=0.7]{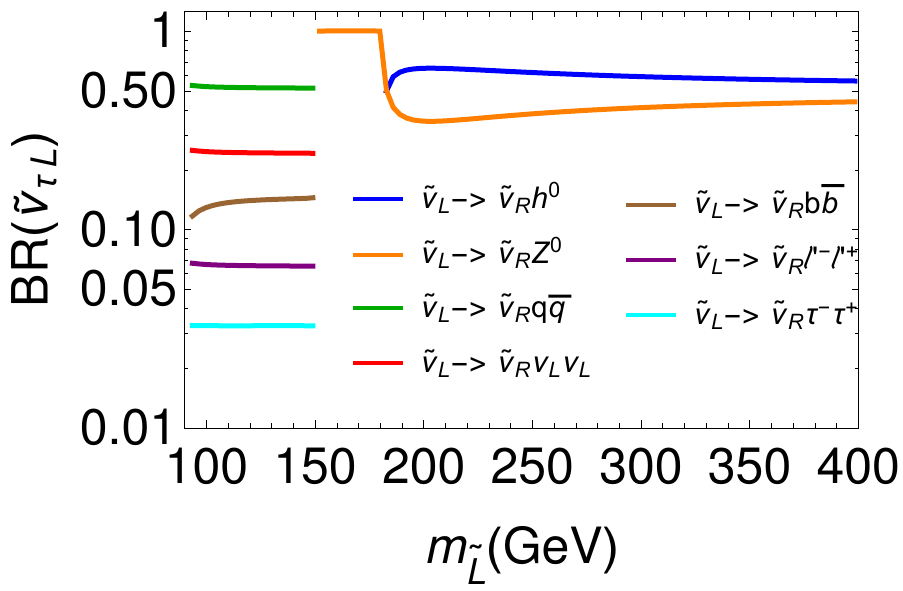} \\
\includegraphics[scale=0.7]{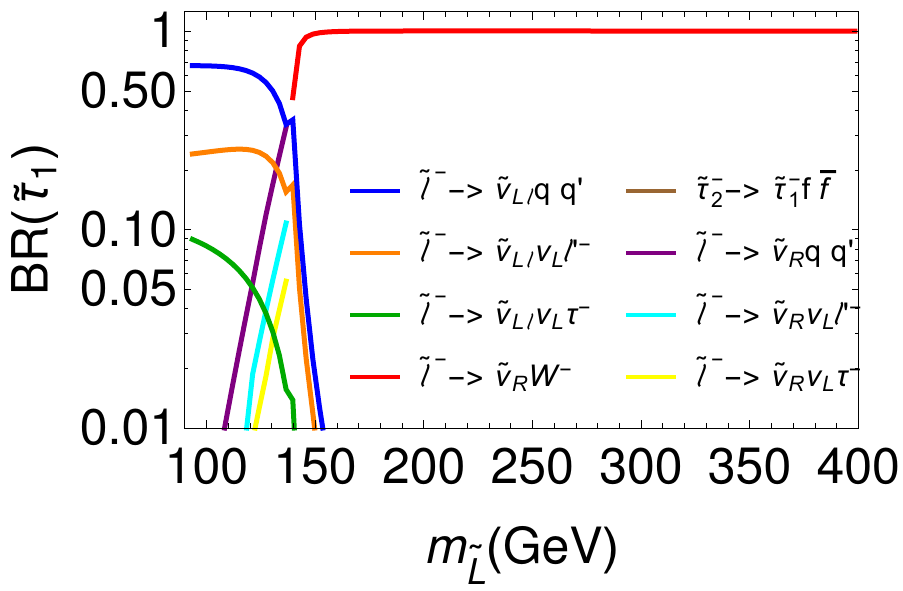} \quad \hspace{177pt}
\caption{\label{fig:SleptonBR2} Branching ratios for sleptons as a function of slepton soft mass $m_{\tilde L}$, for $m_{\tilde L}<\mu=400$~GeV. Decays for charged sleptons (sneutrinos) are shown on the left (right) column. The last panels describe the colours for the branching ratios shown in each column.}
\end{figure*}

The situation changes drastically once the $\mu$ parameter is larger than the slepton mass. In that case, 
the previous decays are not possible, and one either needs to consider alternative two-body channels, 
or new three-body decays. We show the available branching ratios in Figure~\ref{fig:SleptonBR2}, as a function of the slepton mass,
where we have fixed $\mu=400$~GeV.

For $\tilde e_L^-$ and $\tilde \mu_L^-$, we find that the dominant decay is $\tilde\ell^-_L\to\tilde\nu_{\ell L}W^{-*}$, 
with the virtual $W^-$ giving jet pairs or a charged lepton plus a light neutrino. As usual, decays with quark final states have larger branching ratios.

Another possibility is to decay directly to an R-sneutri-no and a real $W^-$ ($\tilde\ell^-_L\to\tilde\nu_{2,3}W^-$). This process depends on the small $LR$ mixing
in the sneutrino sector, so it is proportional to $Y_\nu$. We find that the branching ratio for $\tilde\mu_L^-$ is generally smaller than $20\%$. As mentioned earlier, the $Z_e^{\rm NH}$ factor in $Y_\nu$ is slightly suppressed with respect to $Z_\mu^{\rm NH}$,
so for $\tilde e_L^-$ the branching ratio is smaller.

The stau sector has a slightly different phenomenology, due to the large left-right mixing. In particular, this leads the predictions of this scenario to depend strongly on $\tan\beta$. The mixing splits the states, such that $\tilde\tau_2^-\to\tilde\tau_1^-Z^{0*}$ decay is allowed. The inclusion of this new channel modifies the other $\tilde\nu_{\tau L}\,W^{-*}$ and $\tilde\nu_{2,3}\,W^-$ branching ratios.

The $\tilde\tau_1^-$, on the other hand, for small $m_{\tilde L}$, has similar $\tilde\nu_{\tau L}W^{-*}$ decays, with non negligible $\tilde\nu_{R}W^{-*}$ contributions. The reason for this is that the mixing-induced mass shift implies that $\tilde\tau_1^-$ is close in mass to the $\tilde\nu_{\tau L}$, leading to a strong kinematical suppression.
As a consequence the two-body decay into $\tilde \nu_R\,W^-$ clearly dominates once kinematically allowed,
despite the fact that there is only a small left-right mixing in the sneutrino sector.
Thus, with the exception of $\tilde\tau_1^-$, charged slepton decay shall usually produce one additional $\tilde\nu_{\ell L}$. These states shall be accompanied by jets more than $50\%$ of the time.

We find that all $\tilde\nu_L$ flavours have the same behaviour. For low masses, the decays are governed by 
off-shell $Z^0$- and $h^0$-boson exchange, and a $\tilde\nu_R$ emission.
For larger masses, the branching ratios are dominated by two-body decays into a $\tilde\nu_R$ and an on-shell $Z^0$ or $h^0$ boson, if kinematically allowed.
When both bosons are accessible, the decays into the light Higgs have larger branching ratios.

\subsection{Higgsino LSP ($\mu<m_{\tilde\nu_R}$)}

For completeness, we also study the case where the higgsino is the LSP. This is motivated by the fact that it has not been considered so far in the literature.

In this scenario the sleptons and sneutrinos have two-body decays only as described above for the case of $M_1=M_2=2~{\rm TeV}$. 
We note that, at the one-loop level, a mass splitting between the lightest neutralino and the chargino is induced
via the photon-loop yielding the contribution \cite{Barducci:2015ffa}
\begin{equation}
\Delta m_{\charp} = \frac{|\mu| \alpha(m_Z)}{\pi} \left(2 + \log\left(\frac{|\mu|^2}{m^2_Z}\right) \right)
\end{equation}
This implies that the chargino will always have a sufficiently large decay width such that it decays inside the detector.
However, due to the small mass differences, the decay products of the lightest chargino and the second heaviest neutralino are
so soft that they mainly contribute to the missing transverse momentum.
Note that, in this case, $\tilde \nu_i$ are
hardly produced in the decays of the sleptons and heavier L-sneutrinos.

\section{Set-up}
\label{sec:scan}

For this investigation we have used a series of public programs: 
As a first step we have used \texttt{SARAH} 
\cite{Staub:2008uz,Staub:2013tta,Staub:2012pb,Staub:2010jh,Staub:2009bi} in the 
\texttt{SUSY/BSM toolbox 2.0.1} \cite{Staub:2011dp,Staub:2015kfa} to implement 
the 
model into the event generator \texttt{MadGraph5\_aMC@NLO 2.5.2} 
\cite{Alwall:2014hca}.
For each set of parameters, we use \texttt{SPheno 3.3.8}~\cite{Porod:2003um,Porod:2011nf}
to numerically calculate the mass spectrum and branching ratios.
After generating the hard scattering in \texttt{MadGraph}, the showering and hadronization is carried out internally
with \texttt{PYTHIA 8.233} \cite{Sjostrand:2006za}, which uses the \texttt{CTEQ6L1} PDF set \cite{Pumplin:2002vw}.
We also use \texttt{PYTHIA} for the heavy neutrino decays.
As default we generate  25000 events for every production process.
The generated events are then fed into \texttt{CheckMATE 2.0.7}~\cite{Drees:2013wra,Dercks:2016npn}, which uses \texttt{Delphes 3.4.0} as detector simulation~\cite{deFavereau:2013fsa}.

Given a specific experimental search, \texttt{CheckMATE} compares the number of events passing each signal region 
with the observed $S95$ limit obtained by the experiment via 
the parameter
\begin{align}
\label{eq:r-value-definition}
r^c = \frac{S-1.64\cdot\Delta S}{S^{95}_{obs}}
\end{align}
with $S$ being the number of events in the considered signal region, $\Delta S$ the error from the Monte 
Carlo and $S^{95}_{obs}$ is the experimentally observed 95\% confidence limit on the signal 
\cite{Drees:2013wra,Drees:2015aeo}. In our work, we indicate \texttt{CheckMATE} to compare our predicted signal
with all of the available experimental searches.

Notice that Eq.~(\ref{eq:r-value-definition}) does not capture all theoretical uncertainties, such as missing
higher order corrections in the production and decays of the various particles. Moreover, there are
effects due to variations of the input parameters. For example,
in case of an $\tilde \nu_R$-LSP, the charginos will decay into either a $\mu$ or $\tau$ plus one of
the $R$-sneutrinos which escapes detection. The ratio of $\mu$ over $\tau$ depends on $Y_\nu$ and
varies with the choice of neutrino mixing angles.  
We therefore follow the basic idea presented in ref.\
\cite{Drees:2015aeo} to capture such uncertainties: we do not take the $r^c=1$ value as sharp boundary but
assume that all points with $r^c\ge 1.2$ ($r^c\le 0.8$) are excluded (allowed) whereas for the range in between
one would need a more detailed investigation. 

For each point, we have also checked that the Z and Higgs invisible width respect experimental bounds, and 
that the heavy neutrino mixing is small enough to avoid direct detection~\cite{Deppisch:2015qwa}. 
We have also checked that LFV processes such as $\mu\to e \gamma$, do not exceed the current constraints~\cite{Abada:2014kba}. In this scenario, the SUSY
contribution to LFV is very small, either due to the heavy gaugino masses, or due to the small Yukawa couplings. Thus, the non-SUSY part dominates,
and as one can see in~\cite{Gago:2015vma}, for the current choice of $M_5$, $M_6$ and $\gamma_{56}$, it does not saturate the bounds.

\section{Results}
\label{sec:results}

\subsection{Higgsino LSP}

We study first the case of a higgsino LSP. In these scenarios the sleptons decay directly into
either a lepton and missing energy, or invisibly.
The latter case occurs in case of
$\tilde l \to \nu \charm$ because the decay products of the charginos are very soft.

For the following
investigation we have fixed $M_1=M_2=1$~TeV implying that (i) the $\tilde e$ decays are mainly via
the small gaugino admixtures in the chargino and neutralinos and (ii) there will be practically no
right-handed neutrinos produced in the slepton decays. From this point of view we are effectively
in the usual MSSM with a higgsino LSP. However, to our knowledge the bounds on the slepton mass
parameters due to the LHC data have
not been presented in the literature.
\begin{figure}[tbp]
\begin{center}
\includegraphics[width=0.48\textwidth]{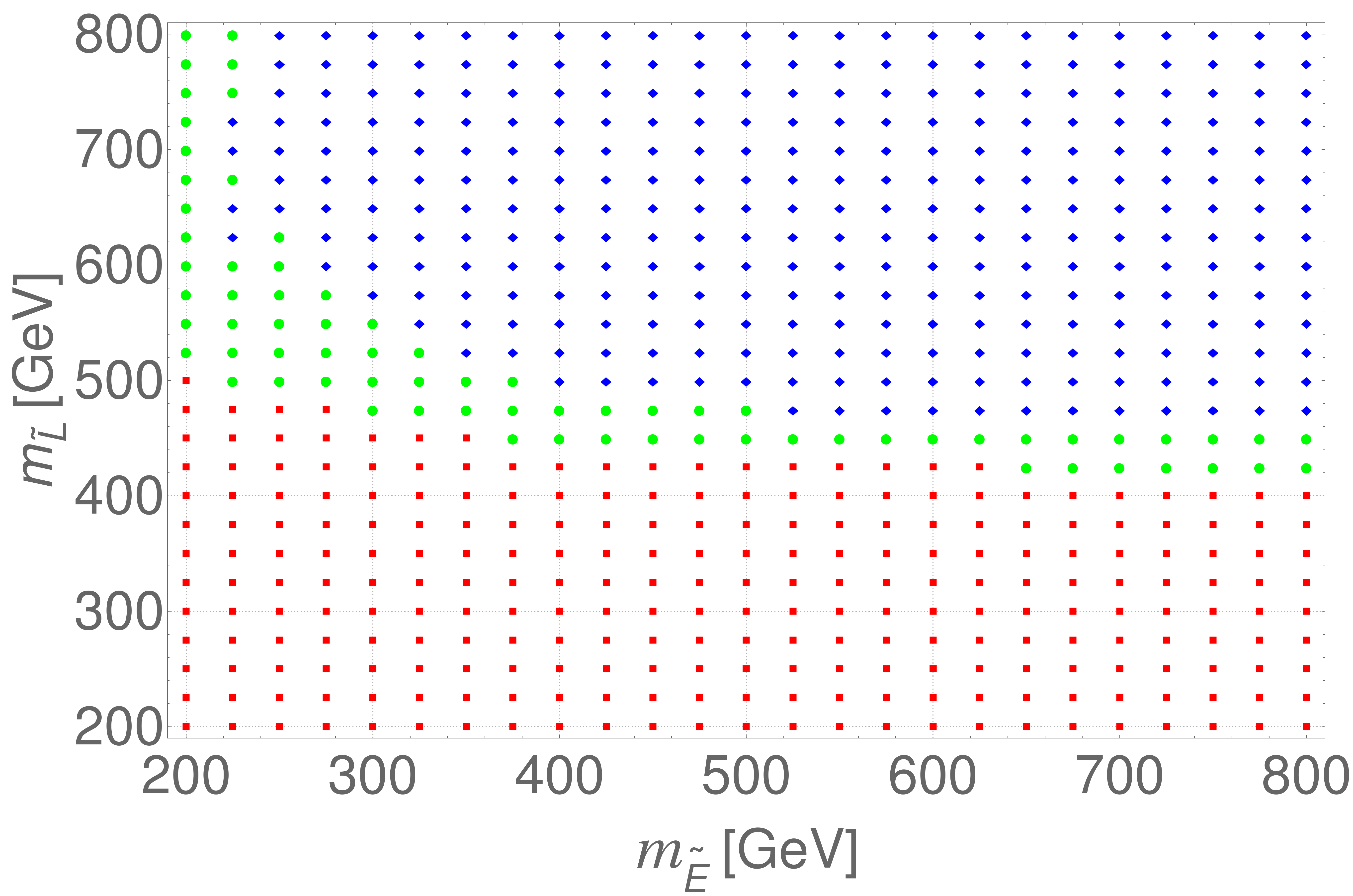}
\caption{\label{fig:MSL_MSE_exclusion} Constraints on combinations of $m_{\tilde E}$
and $m_{\tilde L}$ due to slepton/sneutrino production in case of a higgsino LSP with
$M_1=M_2=1$~TeV, $\mu=120$~GeV  and $\tan\beta=10$. Red points are excluded, blue ones are allowed and
in case of the green ones no conclusive statement can be drawn, within the known 
theoretical and experimental uncertainties.
}
\end{center}
\end{figure}

In this scenario, the most important
analysis is the search for two same sign leptons in combination with large missing transverse 
energy, carried out in~\cite{ATLAS:2016uwq}. This leads to bounds on the $m_{\tilde E}-m_{\tilde L}$ plane,
which are shown in Figure~\ref{fig:MSL_MSE_exclusion} 
for the case $\mu=120$~GeV  and $\tan\beta=10$.
On this Figure, one can see that $m_{\tilde L}<400$~GeV 
is excluded, independent of $m_{\tilde E}$. This constraint increases up to 500~GeV
if, in addition, light $\tilde \ell_R^-$ are present. 

In contrast, even $\tilde \ell_R$ with a mass of 200 GeV cannot be excluded, which can be seen in the Figure in the limit of heavy $\tilde \ell^-_L$.
We understand this is due to insufficient LHC data having been analysed. 
However, this might change in the near future, once
the full 2016 data set has been investigated by ATLAS and CMS.

The structure for $m_{\tilde L}\gsim$~600~GeV and
$m_{\tilde E}\lsim 250$~GeV can be understood from the interplay of different
signal regions defined in~\cite{ATLAS:2016uwq}. These regions differ mainly in the
required bound on the `stransverse' mass $m_{T2}$~\cite{Lester:1999tx,Barr:2003rg}:
$m_{T2} \ge 90$, 120, 150~GeV, corresponding to the signal regions 2LASF, 2LBSF and 2LCSF, respectively. 
Taking, for example, $m_{\tilde{L}} = 625$~GeV, one finds that for $m_{\tilde{E}}  = 200$, 225, 250 and 275 GeV, 
the signal region 2LASF, 2LBSF, 2LBSF and 2LCSF is the most important one, respectively.

Last but not least, we remind that if the mass difference between the sleptons and the higgsinos gets 
too small, then the average value of the transverse moment of the lepton could be below 20 GeV. This can be a problem,
as the $p_T$ cut for the leading (subleading) lepton in this search is of 25~GeV (20~GeV).
In this situation, one cannot carry out any exclusions, as the final states 
are not energetic enough to pass the triggers.

\begin{figure}[tbp]
\begin{center}
\includegraphics[width=0.48\textwidth]{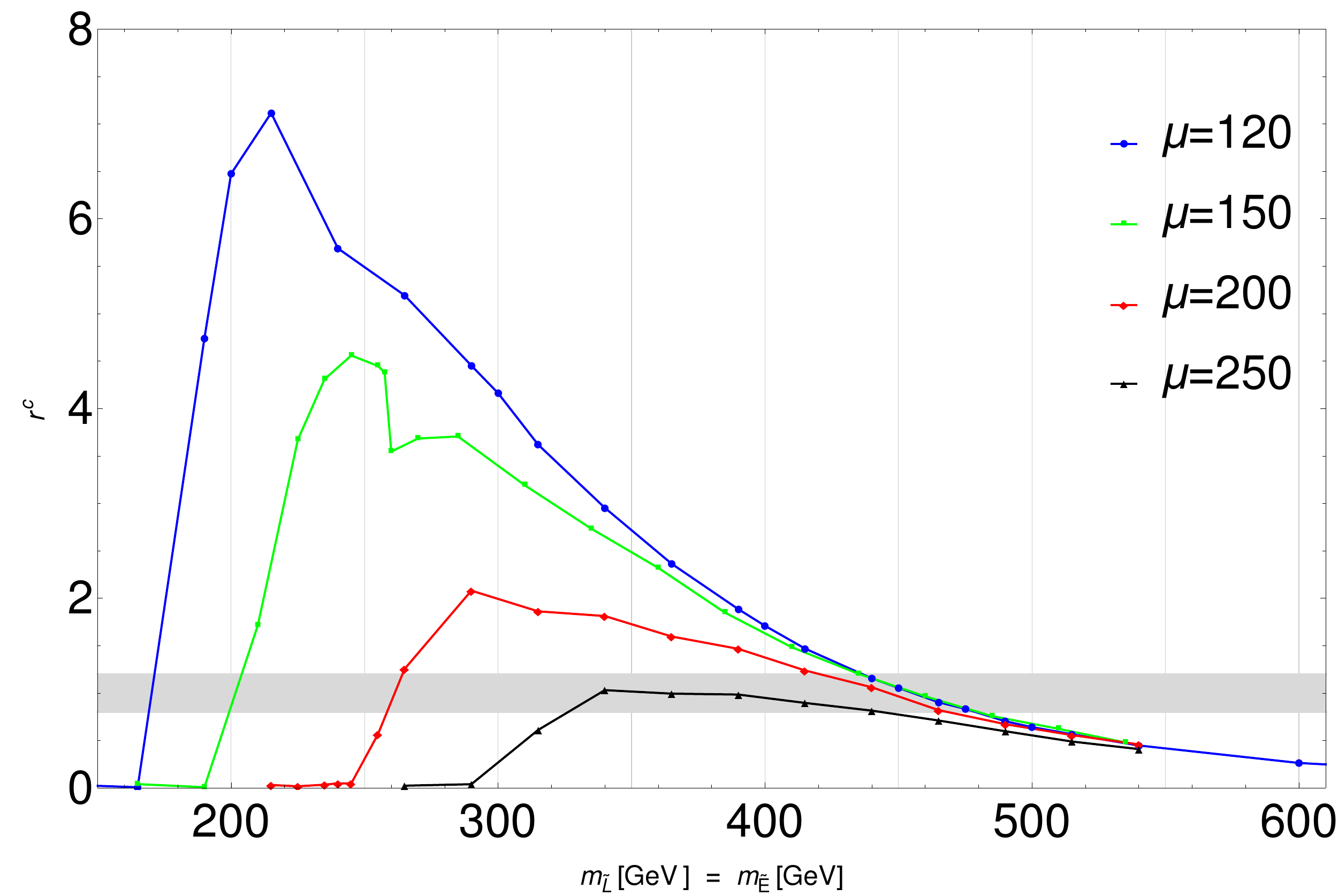}
\caption{\label{fig:r_diagonal} $r^c$ as a function for $\tan\beta=10$, $M_1=M_2=1$~TeV and 
$\mu=120$, 150, 200 and 250~GeV, respectively. The grey band ($0.8\le r^ c\le 1.2$) gives the region
where one cannot draw a conclusion whether the point is allowed or not, values below
are allowed and those above are excluded.}
\end{center}
\end{figure}

We note that the results hardly depend on the value of $\tan\beta$, whose main effect is to enlarge
the mass splitting of the staus for growing values. More important is the size of $\mu$ as this
affects the kinematics, e.g.\ larger values of $|\mu|$ imply softer leptons for fixed slepton mass parameters.
This is demonstrated in \fig{fig:r_diagonal}, where we display the $r^c$-value for different values of $\mu$ as a function of $m_{\tilde L}=m_{\tilde E}$.
As we have mentioned previously, scenarios with
$r^c$ values below 0.8 are allowed, the ones with $r^c>1.2$ are excluded whereas for those in between (gray band)
no conclusive statement can be made. The structure close to the maxima  of the different curves is again
due to the interplay of the different signal regions.

\subsection{R-Sneutrino LSP}

As we have seen in Sections~\ref{sec:snuLSP.mulight} and~\ref{sec:snuLSP.muheavy}, on the R-sneutrino LSP scenario, different decays occur depending on the size of $\mu$ with 
respect to $m_{\tilde L}$. Thus, in order to study this situation appropriately, we first need to understand the 
constraints on chargino pair production.

As mentioned previously, after production, each char-gino decays into a $\tilde\nu_{2,3}$ and a charged 
lepton. Thus, the main constraints arise from the search for two leptons plus missing transverse energy at 13 TeV
\cite{ATLAS:2016uwq}. For very small values of $\mu$, additional constraints arise from the measurement
of the $W^+ W^-$ cross section at 8 TeV, with subsequent decays of the $W$ into leptons 
\cite{ATLAS-CONF-2014-033}. In this case, the charginos would contribute more than what is allowed by the experimental uncertainty
to the $W^+ W^-$ signal regions.

\begin{figure}[tbp]
\centering
\includegraphics[width=0.48\textwidth]{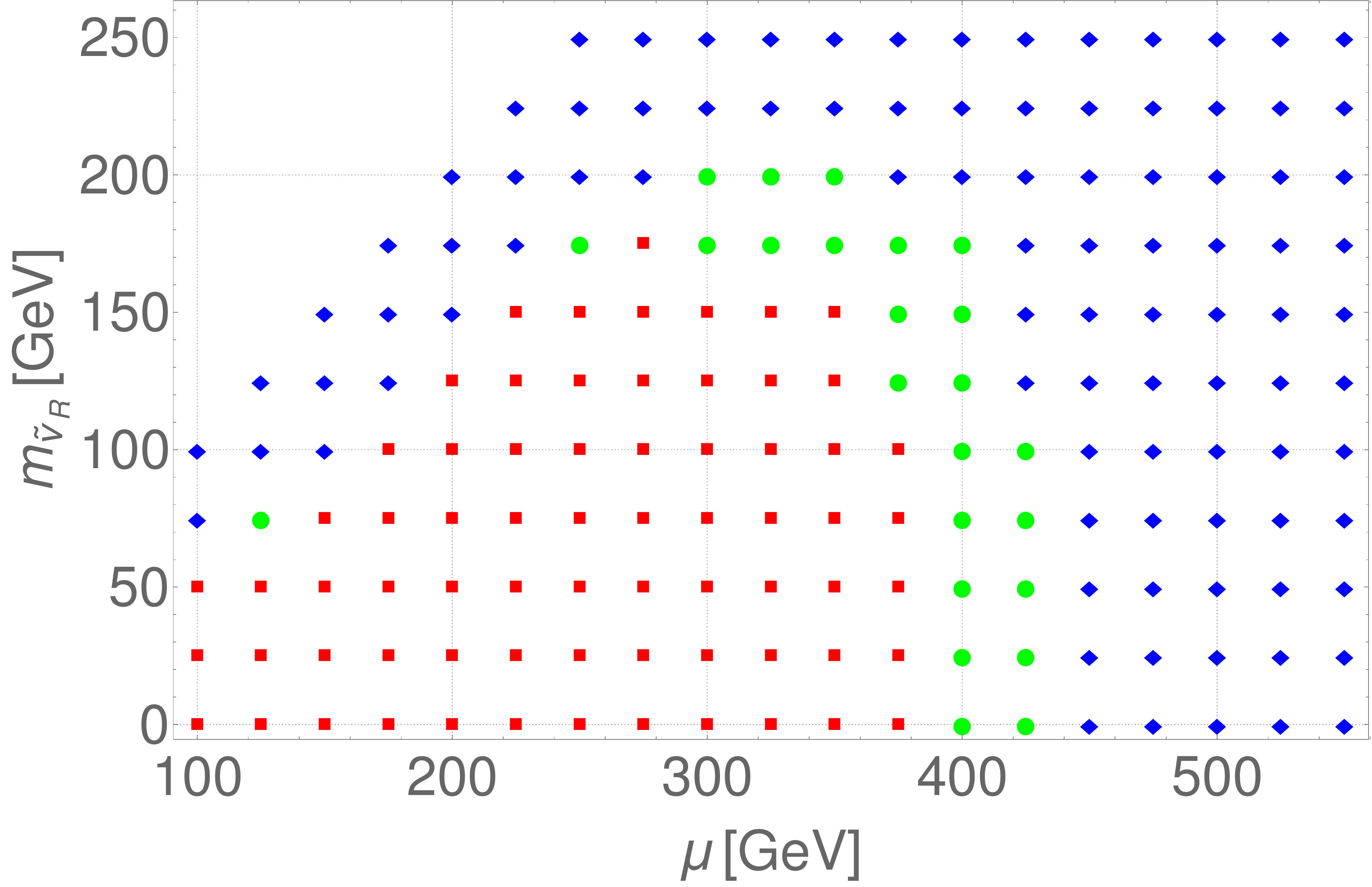}
\caption{\label{fig:chargino_exclusion} Constraints on combinations of $m_{\tilde \nu_R}$ and $\mu$
due to chargino pair production $pp\to \charp\charm \to l^+ l^- \tilde\nu_R\tilde\nu_R^*$. Colour conventions follow Figure~\ref{fig:MSL_MSE_exclusion}.}
\end{figure}
In Figure~\ref{fig:chargino_exclusion}, we show the exclusion region in the $\mu-m_{\tilde\nu_R}$ parameter space, based on 
$\tilde\chi^+\tilde\chi^-$ production.
From the plot, we see that, for vanishing $m_{\tilde\nu_R}$, the bound on $m_{\tilde\chi^\pm}$ can be as large as 375 GeV. 
Note that the $\tilde\nu_{2,3}$, for  $m_{{\tilde\nu}_R}=0$, have nearly the same masses as
the right-handed neutrinos.
For relatively small values of $\mu$, one finds that R-sneutrino masses lighter than $\mu-75$ GeV are ruled out, with the allowed region 
increasing for $\mu\gtrsim250$ GeV.
For completeness, we note that this exclusion does not depend on the sign of $\mu$ or the value $\tan\beta$.

In Section~\ref{sec:snuLSP.mulight}, we analyzed two scenarios for the gauginos, one where $M_1=M_2=2$~TeV, and another ``decoupled'' scenario, where we set $M_1=M_2=1$~PeV. Given our results for chargino production, we explore two additional possibilities. On the first one (``varying $\mu$''), we set $\mu=m_{\tilde\nu_R}+25$~GeV, such that we always have $m_{\tilde\nu_R}<\mu<m_{\tilde L}$. On the second one (``fixed $\mu$''), we set $\mu=400$~GeV, such that one also needs to take into account the $m_{\tilde\nu_R}<m_{\tilde L}<\mu$ case. Thus, four different exclusion plots will be generated. In all of these, we merge exclusions from 8 and 13 TeV data.

\begin{figure*}
\centering
\includegraphics[width=0.48\textwidth]{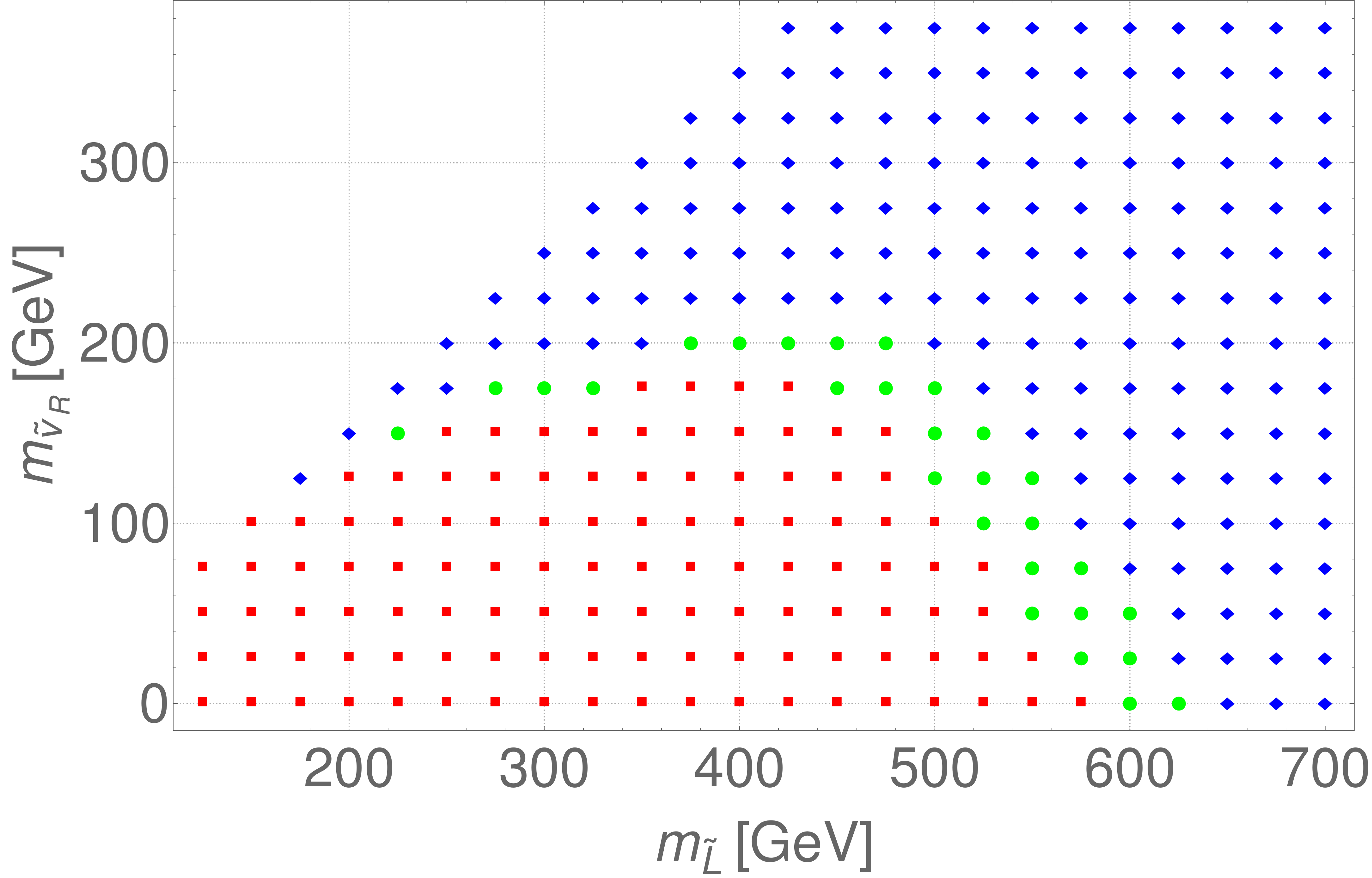} \hfill
\includegraphics[width=0.48\textwidth]{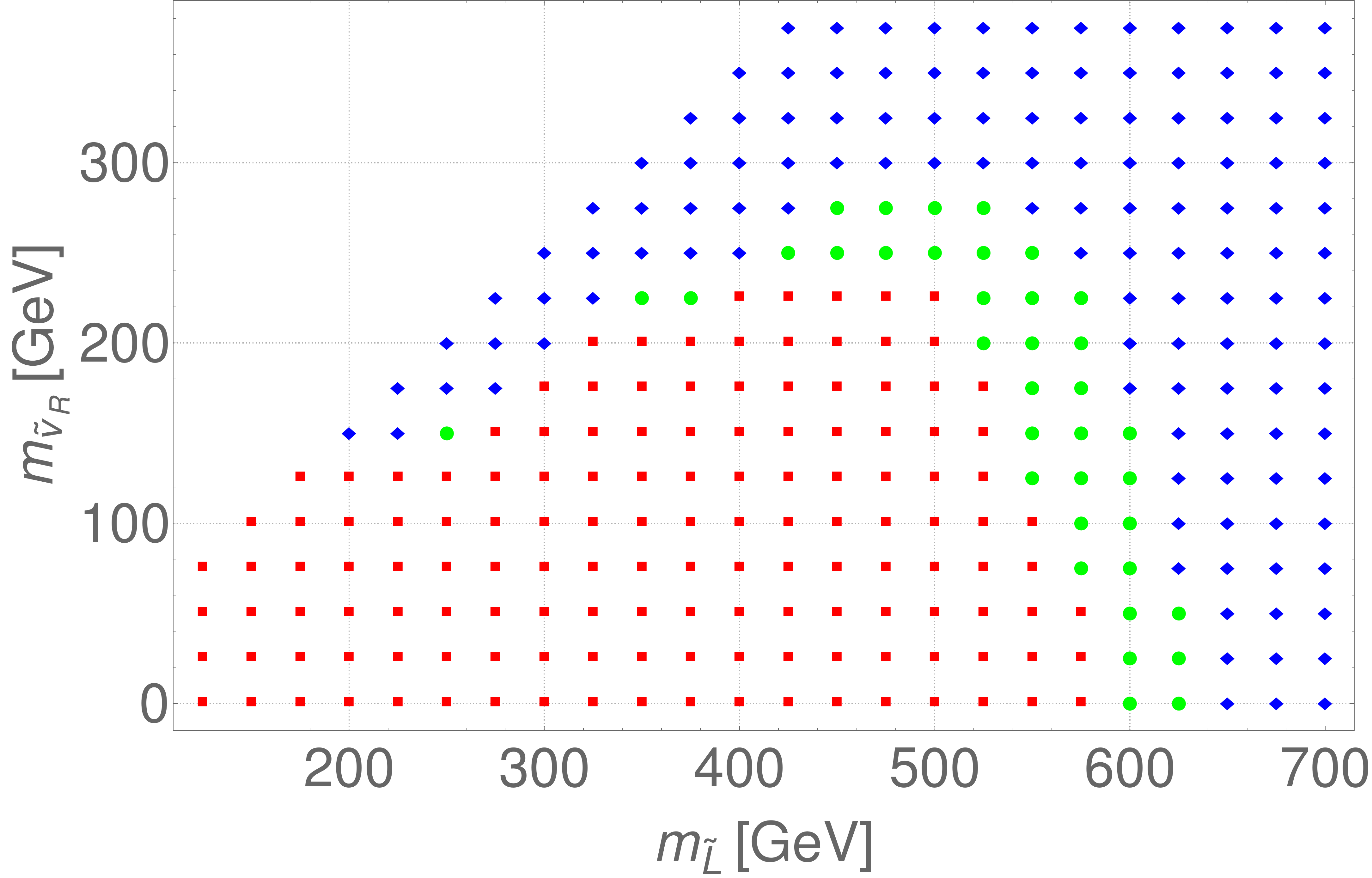}
\caption{\label{fig:varymu-exclusion} Constraints on combinations of $m_{\tilde \nu_R}$
and $m_{\tilde L}$ due to slepton/sneutrino production in case of an R-sneutrino LSP with
$M_1=M_2=2$~TeV ($M_1=M_2=1$~PeV) on the left (right) panel. We fix $\mu=m_{\tilde\nu_R}+25$~GeV  and $\tan\beta=6$. Colour conventions follow Figure~\ref{fig:MSL_MSE_exclusion}.}
\end{figure*}
We show the constraints on the varying $\mu$ scenario in Figure~\ref{fig:varymu-exclusion}. Here, the relevant analysis is again~\cite{ATLAS:2016uwq}, which searches for events with 2-3 leptons and missing energy. We find that, for both choices of gaugino mass, we can rule out values of $m_{\tilde L}$ as large as 575~GeV. In addition, for lighter slepton masses, it is possible to rule out R-sneutrino masses as heavy as 175-225 GeV, depending on the amount of gaugino admixture.

The exclusion for ``decoupled'' gauginos is stronger, which can be understood from Figure~\ref{fig:SleptonBR1}. For $\tilde\ell_L$, all possible decays shall lead at least to one charged lepton, for all values of gaugino mass (recall that, in this scenario, charginos decay always into final states with visible charged leptons). However, for $\tilde\nu_{\ell L}$, one finds that it is possible to have only missing energy on the final state, due to $\tilde\nu_{\ell L}\to\nu_L\tilde\chi^0$ decay. This decay channel is suppressed in the ``decoupled'' scenario, meaning that it is much more likely to have energetic charged leptons on the final state, which strengthens the multi-lepton signal.

\begin{figure*}
\centering
\includegraphics[width=0.48\textwidth]{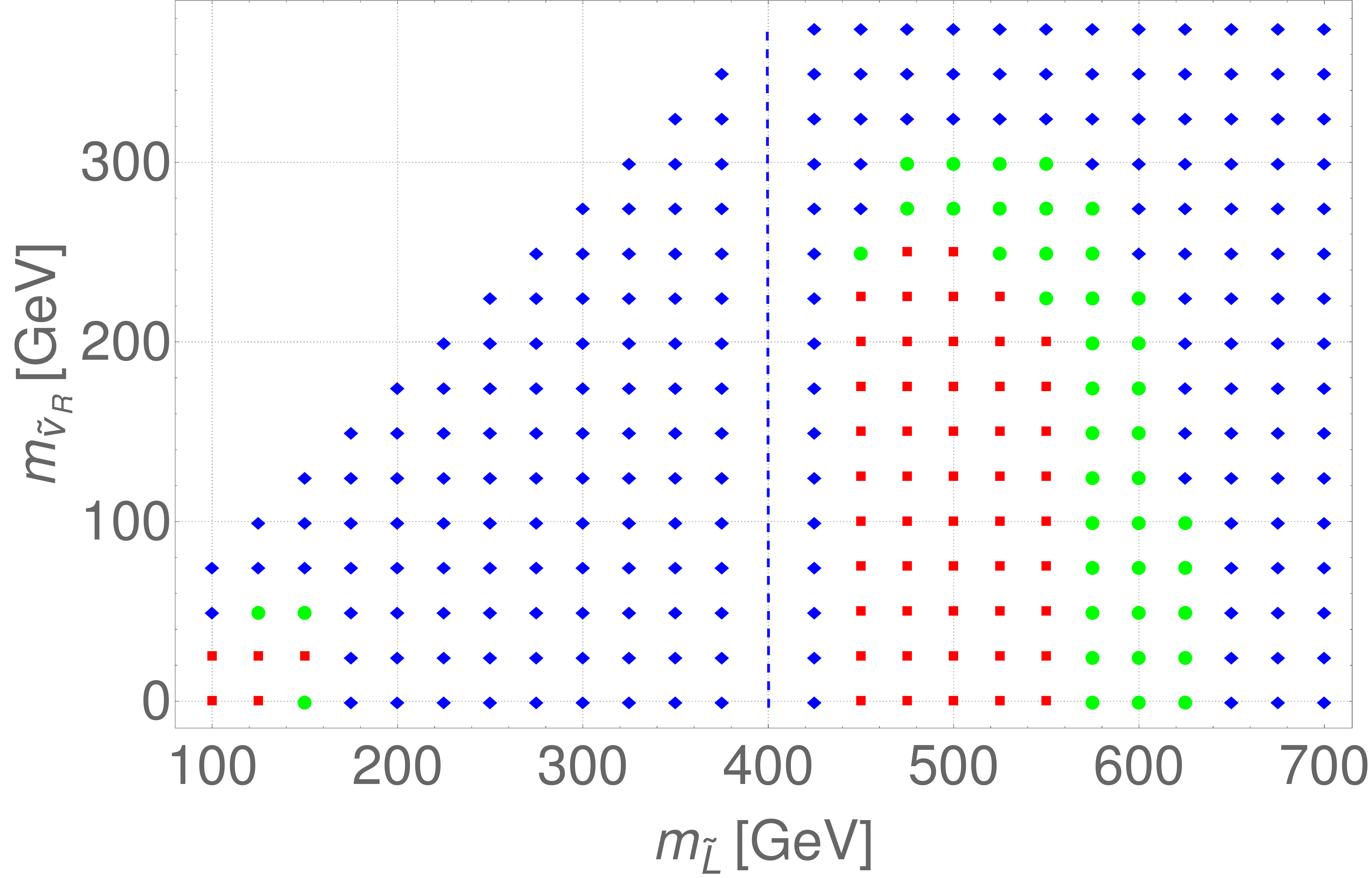} \hfill
\includegraphics[width=0.48\textwidth]{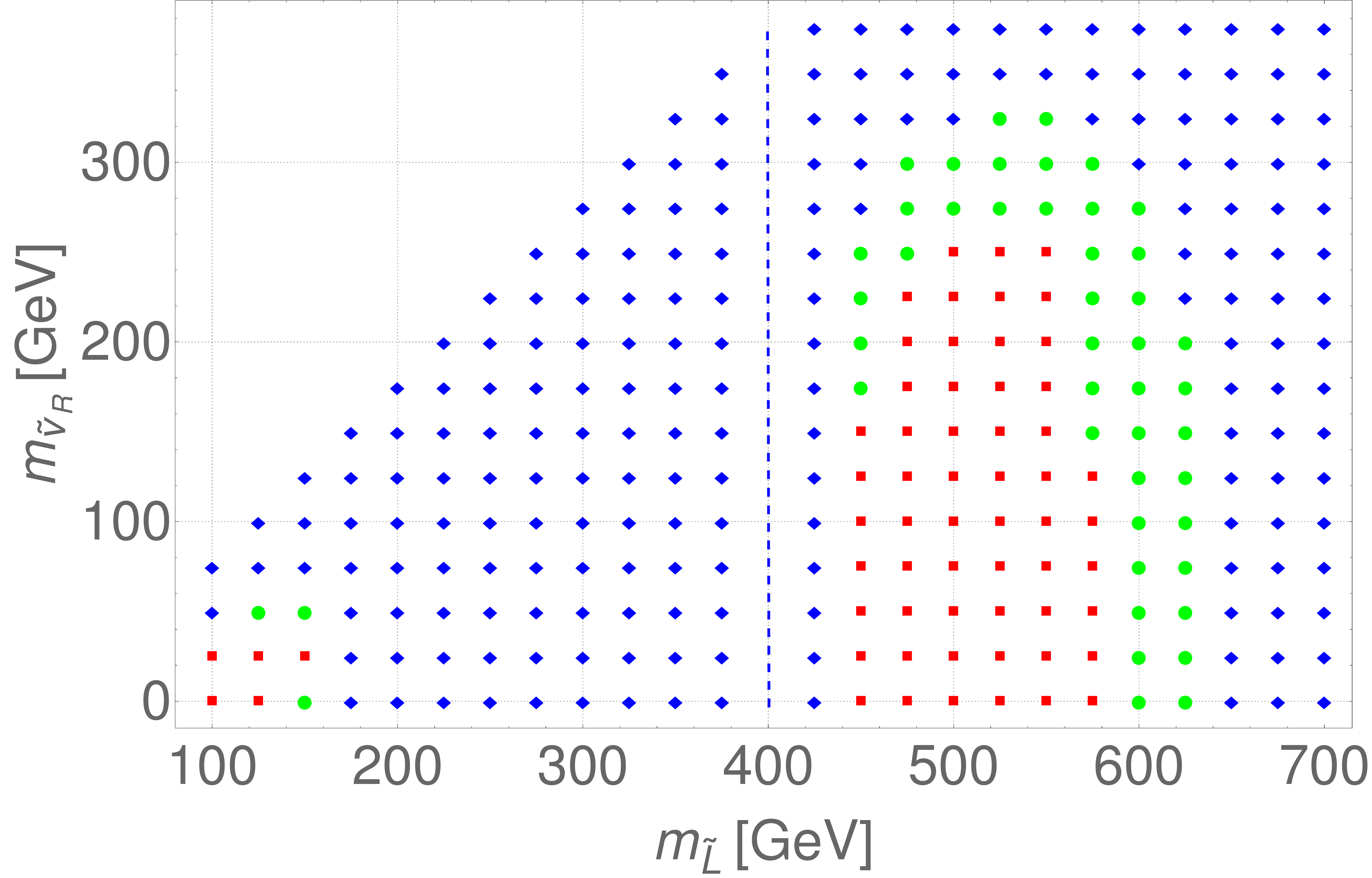}
\caption{\label{fig:fixedmu-exclusion} Constraints on combinations of $m_{\tilde \nu_R}$
and $m_{\tilde L}$ due to slepton/sneutrino production in case of an R-sneutrino LSP with
$M_1=M_2=2$~TeV ($M_1=M_2=1$~PeV) on the left (right) panel. We fix $\mu=400$~GeV  and $\tan\beta=6$. Colour conventions follow Figure~\ref{fig:MSL_MSE_exclusion}.}
\end{figure*}
The constraints on the fixed $\mu$ case are shown in Figure~\ref{fig:fixedmu-exclusion}, again, for different values of gaugino mass. Here it is very interesting to note that there are very weak constraints when $m_{\tilde L}<\mu$. The reason for this can be found in Figure~\ref{fig:SleptonBR2}. We see that in most of the cases, we have the charged slepton decaying into $\tilde\nu_{\ell L}$ and light fermions. Given the proximity in the slepton masses, most light fermions end up being very soft, and elude detection. On the other hand, when the two resulting $\tilde\nu_{\ell L}$ decay, the decay products shall involve either an on-shell $h^0$ or $Z^0$. This is again problematic for detection, as fermions coming from these states are generally avoided in new physics searches by suitable cuts to suppress SM background. This leaves us sensitive only to the very low $m_{\tilde L}$ region, where L-sneutrino three-body decays are allowed.

For large values of $m_{\tilde L}$, we return to the $\mu<m_{\tilde L}$ scenario. Here, again, we have~\cite{ATLAS:2016uwq} giving the relevant constraints. The bounds reach $m_{\tilde L}$ as large as 575~GeV for vanishing $m_{\tilde\nu_R}$. Morover, for smaller charged slepton masses, we can bound $m_{\tilde\nu_R}$ up to 250~GeV. This is all consistent with our results for the varying $\mu$ scenario.

\section{Conclusions}
\label{sec:conclusion}

In this work, the MSSM was extended by three right-handed neutrino superfields, with active neutrino masses being provided through  the seesaw mechanism. In addition, driven by naturalness arguments, the $\mu$ term was kept relatively small, such that the lightest neutralinos were higgsino-like. We considered LHC data on this model, and explored how much
existing data constrain such scenarios.

Two possibilities were considered for the nature of the LSP. On the first one, this was a higgsino-like neutralino. In this case, one requires a non-vanishing gaugino admixture in order not to have too long-lived charginos. The sleptons would decay into SM particles and neutralinos. We carried out an analysis considering $m_{\tilde L}\neq m_{\tilde E}$, for fixed neutralino mass, and found that only $m_{\tilde L}$ could be bounded. For $\mu=120$~GeV, we can rule out at least $m_{\tilde L}<400$~GeV for all values of $m_{\tilde E}$, and $m_{\tilde L}<500$~GeV for $m_{\tilde E}=200$~GeV.

On the second possibility we considered, the right-handed neutrino superpartner, the R-sneutrino, was taken as the LSP. This provided a very complex scenario, depending on the relative size of the neutralino / chargino mass with respect to the slepton mass. The phenomenology also depended on the amount of gaugino component within the neutralinos. We found that, as long as $\mu<m_{\tilde L}$, we can exclude slepton masses to a maximum of $575$~GeV, for vanishing $m_{\tilde\nu_R}$. For lower values of slepton mass, the R-sneutrino masses can be excluded up to about 200~GeV. In case $m_{\tilde L}<\mu$, constraints became very weak, as final states were either too soft, or excluded from signal regions.

%

\section*{Acknowledgements}
This work has been supported by BAYLAT, project nr.\ 914-20.1.1.
T.F.\ and W.P.\ have been supported by  the DFG, project nr.\ PO-1337/3-1.
J.J.P.~acknowledges funding by the {\it Direcci\'on de Gesti\'on de la Investigaci\'on} at PUCP, through grant DGI-2015-3-0026. N.C.V. has been supported by CienciActiva-CONCYTEC Grant 026-2015. We also acknowledge the financial support of PUCP DARI Groups Research Fund 2016. We would especially like to thank O.~A.~D\'iaz in the {\it Direcci\'on de Tecnolog\'ias de Informaci\'on} at PUCP, for implementing the code within the {\tt LEGION} system.

\appendix

\section{Parametrization of the Neutrino Sector}
\label{app:numass}

In order to parametrize neutrino mixing, we generalize the work in~\cite{Casas:2001sr,Donini:2012tt} to three heavy neutrinos. The $6\times6$ neutrino mixing matrix $U$ is decomposed into four $3\times3$ blocks:
\begin{equation}
 U_{6\times 6}=\left(\begin{array}{cc}
U_{a\ell} & U_{ah} \\
U_{s\ell} & U_{sh}
\end{array}\right)~.
\end{equation}
where $a=e,\,\mu,\,\tau$ and $s=s_1,\,s_2,\,s_3$ make reference to the active and sterile states, while $\ell=1,\,2,\,3$ and $h=4,\,5,\,6$ refer to the light and heavy mass eigenstates, respectively.

For the normal hierarchy, each block can be parame-trized in the following way:
\begin{align}
 \label{eq:mixingmats}
 U_{a\ell} &= U_{\rm PMNS}\, H~,  \\
 U_{ah} &= i\,U_{\rm PMNS}\, H\,m_\ell^{1/2}R^\dagger M_h^{-1/2}~, \\
 U_{s\ell} &= i\bar H M_h^{-1/2}\,R\,m_\ell^{1/2}~, \\
 U_{sh} &= \bar H~,
\end{align}
where $m_\ell={\rm diag}(m_1,\,m_2,\,m_3)$ and $M_h={\rm diag}(M_4,\,M_5,$ $\,M_6)$ are $3\times3$ matrices including the light and heavy neutrino masses, respectively, and:
\begin{eqnarray}
\label{eq:hreal}
H &=& \left(I+m_\ell^{1/2}\,R^\dagger\,M_h^{-1}\,R\,m_\ell^{1/2}\right)^{-1/2} \nonumber \\
\bar H &=& \left(I+M_h^{-1/2}\,R\,m_\ell\,R^\dagger\,M_h^{-1/2}\right)^{-1/2}~.
\end{eqnarray}
In addition, $U_{\rm PMNS}$ corresponds to the standard PMNS matrix in the limit where $H\to I$, and $R$ is a complex orthogonal matrix as in~\cite{Casas:2001sr}, which we parametrize in the following way:
\begin{equation}
 R=\left(\begin{array}{ccc}
 1 & & \\
 & c_{56} & s_{56} \\
 & -s_{56} & c_{56}
\end{array}\right)\left(\begin{array}{ccc}
c_{46} & & s_{46} \\
& 1 & \\
-s_{46} & & c_{46}
\end{array}\right)\left(\begin{array}{ccc}
c_{45} & s_{45} & \\
-s_{45} & c_{45} & \\
& & 1
\end{array}\right)
\end{equation}
Here, $s_{ij}$ and $c_{ij}$ are, respectively, the sine and cosine of a complex mixing angle, $\rho_{ij}+i\gamma_{ij}$. With these parameters, one can rebuild the $Y_\nu$ and $M_R$ matrices, meaning that the neutrino sector is described without ambiguities:
\begin{eqnarray}
 Y_\nu &=& -i\frac{\sqrt{2}}{v_u}U_{\rm PMNS}^*H^*m_\ell^{1/2}\left(m_\ell R^\dagger+R^TM_h\right)M_h^{-1/2}\bar H \nonumber \\ \\
 M_R &=& \bar H^* M_h^{1/2}\left(I-M_h^{-1}R^*m_\ell^2R^\dagger M_h^{-1}\right)M_h^{1/2}\bar H
\end{eqnarray}

In general, the active-heavy mixing is suppressed by $(m_\ell/M_h)^{1/2}$, which would imply heavy neutrinos being difficult to probe if their masses are much heavier than those of the light neutrinos. However, this result can be avoided by taking large $\gamma_{ij}$. Since these involve hyperbolic sines and cosines, at least one large $\gamma_{ij}$ would lead to an exponential enhancement of the mixing.

In order to simplify our analysis, in the following we keep $\nu_4$ decoupled from $\nu_5$ and $\nu_6$, setting $\rho_{45}=\rho_{46}=\gamma_{45}=\gamma_{46}=0$, that is, only $\rho_{56}$ and $\gamma_{56}$ are not zero. For GeV masses, and $\gamma_{56}$ in the range $3-10$, we find the standard results:
\begin{eqnarray}
\label{eq:dm-mixing}
|U_{a4}|^2&=&\left|(U_{\rm PMNS})_{a1}\right|^2\frac{m_1}{M_4} \\
|U_{a5}|^2&=&\left|Z^{NH}_a\right|^2\frac{m_3}{M_5}\cosh^2\gamma_{56} \\
|U_{a6}|^2&=&\left|Z^{NH}_a\right|^2\frac{m_3}{M_6}\cosh^2\gamma_{56}
\end{eqnarray}
where $Z^{NH}_a$ is a factor of $\ord{1}$ depending on the PMNS mixing angles and the neutrino mass ordering, which can be found in~\cite{Gago:2015vma}. This limit also leads to Eqs.~\ref{eq:YukawasSimple}.

\section{Sneutrino Mass Matrix}
\label{app:snumassmatrix}

Being electrically neutral, the sneutrino interaction states can be split into real and imaginary parts:
\begin{align}
 \tilde\nu_L=\frac{1}{\sqrt{2}}\left(\tilde\phi_{LR}+i\tilde\phi_{LI}\right) &&
 \tilde\nu_R^c = \frac{1}{\sqrt{2}}\left(\tilde\phi_{RR}-i\tilde\phi_{RI}\right)
\end{align}
We can define $\tilde\phi_R=(\phi_{LR},\,\phi_{RR})^T$ and $\tilde\phi_I=(\phi_{LI},\,\phi_{RI})^T$, such that the sneutrino mass term is divided into four $2\times2$ blocks:
\begin{equation}
 \mathcal{L}_{\tilde\nu}^{\rm mass}=\frac{1}{2}(\tilde\phi_R^T,\,\tilde\phi_I^T)\cdot\left(\begin{array}{cc}
M^2_{RR} & M_{RI}^2 \\
M_{IR}^2 & M_{II}^2
\end{array}\right)\cdot\left(\begin{array}{c}
\tilde\phi_R \\ \tilde\phi_I
\end{array}\right)
\end{equation}

\begin{widetext}
The blocks are:
\begin{equation}
 M_{RR}^2=\left(\begin{array}{cc}
m_{\tilde L}^2+\frac{1}{2}m_Z^2\cos2\beta+\frac{1}{2}v_u^2Y_\nu Y_\nu^\dagger & 
\Re e\left[\frac{v_u}{\sqrt{2}}\left(T_\nu+Y_\nu M_R^\dagger-\mu^*Y_\nu\cot\beta\right)\right] \\
\Re e\left[\frac{v_u}{\sqrt{2}}\left(T^T_\nu+M_R^\dagger Y_\nu^T-\mu^*Y^T_\nu\cot\beta\right)\right] & 
m_{\tilde \nu_R}^2+M_R^\dagger M_R+\frac{1}{2}v_u^2Y_\nu^\dagger Y_\nu +\Re e[B_{\tilde \nu}]
\end{array}\right) 
\end{equation}
\begin{equation}
 M_{II}^2=\left(\begin{array}{cc}
m_{\tilde L}^2+\frac{1}{2}m_Z^2\cos2\beta+\frac{1}{2}v_u^2Y_\nu Y_\nu^\dagger & 
\Re e\left[\frac{v_u}{\sqrt{2}}\left(T_\nu-Y_\nu M_R^\dagger-\mu^*Y_\nu\cot\beta\right)\right] \\
\Re e\left[\frac{v_u}{\sqrt{2}}\left(T^T_\nu-M_R^\dagger Y_\nu^T-\mu^*Y^T_\nu\cot\beta\right)\right] & 
m_{\tilde \nu_R}^2+M_R^\dagger M_R+\frac{1}{2}v_u^2Y_\nu^\dagger Y_\nu -\Re e[B_{\tilde \nu}]
\end{array}\right) 
\end{equation}
\begin{equation}
 M_{RI}^2=\left(\begin{array}{cc}
0 & 
\Im m\left[\frac{v_u}{\sqrt{2}}\left(T_\nu+Y_\nu M_R^\dagger+\mu^*Y_\nu\cot\beta\right)\right] \\
-\Im m\left[\frac{v_u}{\sqrt{2}}\left(T^T_\nu+M_R^\dagger Y_\nu^T-\mu^*Y^T_\nu\cot\beta\right)\right] & 
 -\Im m[B_{\tilde \nu}]
\end{array}\right) 
\end{equation}
\begin{equation}
 M_{IR}^2=(M_{RI}^2)^T
\end{equation}
\end{widetext}

It is possible to avoid the splitting of $\tilde \nu_L$ and $\tilde \nu_R^c$ into $\tilde\phi_{(L,R)(R,I)}$ if $M^2_{RR}=M^2_{II}$ and $M^2_{RI}=M^2_{IR}=0$. This is achieved by taking CP conservation, as well as vanishing $B_{\tilde \nu}$ and $Y_\nu M_R^\dagger$. In this work, we have real $\mu$ and $T_\nu$, and very small $Y_\nu M_R^\dagger$ and $B_{\tilde \nu}$. Thus, to a very good approximation, we can assume that the real and imaginary parts of each field are aligned, so we can work directly with $\tilde\nu_L$ and $\tilde\nu_R^c$.

\bibliographystyle{epjc}
\bibliography{Rsneutrino}

\end{document}